\documentclass{svjour3}
\usepackage{graphicx}
\usepackage{float}
\usepackage[justification=centering]{subfig}
\usepackage{cite}
\usepackage{hyperref}
\hypersetup{hidelinks}
\usepackage{url}
\usepackage{listings}
\usepackage{xcolor}
\usepackage{tcolorbox}
\usepackage{caption}
\usepackage{booktabs}
\usepackage{multirow}
\usepackage{enumerate}
\usepackage{textcomp}
\usepackage{booktabs}
\usepackage[sort,compress]{natbib}
\usepackage{diagbox}
\usepackage{enumitem}
\usepackage[misc,geometry]{ifsym}

\newtcolorbox{qoutebox}[3][]
{
  colframe = gray!30!white,
  colback  = #2!10,
  #1,
}

\begin{document}

\title{Code Smells Detection via Modern Code Review: A Study of the OpenStack and Qt Communities}
\titlerunning{Code Smells Detection via Modern Code Review}

\author{Xiaofeng Han    \and
        Amjed Tahir     \and
        Peng Liang      \and
        Steve Counsell  \and
        Kelly Blincoe   \and
        Bing Li         \and
        Yajing Luo
}

\institute{Xiaofeng Han \and Peng Liang (\Letter) \and  Bing Li \and Yajing Luo \at School of Computer Science, Wuhan University, Wuhan, China \\ 
            Hubei Luojia Laboratory, Wuhan, China\\
            \email{\{hanxiaofeng, liangp, bingli, luoyajing\}@whu.edu.cn}
            \and
            Amjed Tahir \at School of Mathematical and Computational Sciences, Massey University, Palmerston North, New Zealand \\ 
            \email{a.tahir@massey.ac.nz}
            \and
            Steve Counsell \at Department of Computer Science, Brunel University London, London, United Kingdom \\ 
            \email{steve.counsell@brunel.ac.uk}
            \and 
            Kelly Blincoe \at Department of Electrical, Computer, and Software Engineering, University of Auckland, Auckland, New Zealand \\ 
            \email{k.blincoe@auckland.ac.nz}
}

\date{Received: date / Accepted: date}

\maketitle

\begin{abstract}
Code review plays an important role in software quality control. A typical review process involves a careful check of a piece of code in an attempt to detect and locate defects and other quality issues/violations. One type of issue that may impact the quality of software is code smells - i.e., bad coding practices that may lead to defects or maintenance issues. Yet, little is known about the extent to which code smells are identified during modern code review. To investigate the concept behind code smells identified in modern code review and what actions reviewers suggest and developers take in response to the identified smells, we conducted an empirical study of code smells in code reviews by analyzing reviews from four large open source projects from the OpenStack (Nova and Neutron) and Qt (Qt Base and Qt Creator) communities. We manually checked a total of 25,415 code review comments obtained by keywords search and random selection; this resulted in the identification of 1,539 smell-related reviews which then allowed the study of the causes of code smells, actions taken against identified smells, time taken to fix identified smells, and reasons why developers ignored fixing identified smells. Our analysis found that 1) code smells were not commonly identified in code reviews, 2) smells were usually caused by \textit{violation of coding conventions}, 3) reviewers usually provided constructive feedback, including fixing (refactoring) recommendations to help developers remove smells, 4) developers generally followed those recommendations and actioned the changes, 5) once identified by reviewers, it usually takes developers less than one week to fix the smells, and 6) the main reason why developers chose to ignore the identified smells is that it is \emph{not worth fixing the smell}. Our results suggest the following: 1) developers should closely follow coding conventions in their projects to avoid introducing code smells, 2) \textit{review-based} detection of code smells is perceived to be a trustworthy approach by developers, mainly because reviews are context-sensitive (as reviewers are more aware of the context of the code given that they are part of the project's development team), and 3) program context needs to be fully considered in order to make a decision of whether to fix the identified code smell immediately.

\end{abstract}

\section{Introduction}
\label{sec:introduction}

Code smells are defined as symptoms of possible code or design problems \citep{martin1999refactoring}, which may potentially have a negative impact on software quality, such as maintainability \citep{palomba2018@maintainability}, code readability \citep{Abbes2011}, testability \citep{Tahir2016}, and defect-proneness \citep{Khomh2009}.

A large number of studies have focused on smell detection and removal techniques \citep{Tsantalis2009,moha2009decor}. There are also a number of open source (and widely used in industrial settings) static analysis tools for smell detection; these include tools such as PMD\footnote{\url{https://pmd.github.io}}, SonarQube\footnote{\url{https://www.sonarqube.org}}, and Designite\footnote{\url{https://www.designite-tools.com}}. 
Those tools are general purpose and use a threshold approach for identifying certain smells. However, previous work \citep{Yamashita2013e,tahir2020stackexchange} has shown that the program context and domain are important in identifying smells. This may also include other factors like developer experience and previous involvement in the project, which makes it difficult for program analysis tools to correctly identify smells since this information is rarely taken into account. Existing smell detection tools are also known to produce false positives \citep{Fontana2016,sharma2018survey}. Therefore, manual detection of smells could be considered more valuable than current automatic approaches.

Code review is a process which aims to verify the quality of software by detecting defects and other issues in the code and to ensure that the code is readable, understandable and maintainable before they are merged into the code base. It has been linked to improved quality \citep{baker1997code}, reduced defects \citep{mcintosh2016empirical}, reduced anti-patterns \citep{morales2015code} and the identification of vulnerabilities \citep{meneely2014empirical}. Compared to smell detection tools, code reviews are usually performed by developers belonging to the same project \citep{mcconnell2004code}, so it is possible that reviewers will take full account of program context and thus better identify code smells. In modern code review (MCR), code changes are reviewed through some code review platforms, such as Gerrit\footnote{\url{https://www.gerritcodereview.com}}. There is a common practice of keeping the changes as small as possible to facilitate the review. But it is still important to see if reviewers identify any code smells in the changes being added to the code base. We are interested in this human-driven detection of code smells since we want to see how different this is from automatic detection of code smells given that context matters. However, little is known about the extent to which code smells are identified during modern code review and whether developers (the code authors) take any action when a piece of code is deemed ``smelly'' by reviewers.

Therefore, we set out to study the concept behind code smells identified in MCR and track down actions taken after reviews were carried out. To this end, we mined code review discussions from four most active projects from the OpenStack\footnote{\url{https://www.openstack.org}} community (Nova\footnote{\url{https://wiki.openstack.org/wiki/Nova}} and Neutron\footnote{\url{https://wiki.openstack.org/wiki/Neutron}}) and the Qt\footnote{\url{https://www.qt.io/}} community (Qt Base\footnote{\url{https://github.com/qt/qtbase}} and Qt Creator\footnote{\url{https://www.qt.io/product/development-tools}}), which use Gerrit as their code review platform. We then conducted a comprehensive quantitative and qualitative analysis to study how common it was for reviewers to identify code smells during code review, why the code smells were introduced, what actions they recommended for those smells, how long it took developers to change the code based on those recommendations and why developers ignored some of the identified smells. In total, we analysed 1,539 smell-related code reviews obtained by manually checking 25,415 review comments to achieve our goal. Our results suggest that: 

\begin{enumerate}[itemindent=1.5em]
    \item Code smells are not widely identified in modern code review.
    \item Following coding conventions can help reduce the introduction of code smells.
    \item Reviewers usually provide useful suggestions to help developers better fix the identified smells, while developers commonly accept reviewer recommendations regarding the identified smells and tend to refactor their code based on those recommendations. 
    \item Review-based detection of code smells is seen as a trustworthy mechanism by developers.
    \item Program context needs to be taken into full account to determine whether to fix the identified code smells immediately.
\end{enumerate}

In this paper, we extended our earlier work on studying code smells in code reviews \citep{xiaofeng2021Understanding} through the following additions:

\begin{enumerate}[itemindent=1.5em]
    \item We extended our dataset by including the code review data from two large projects of the Qt community.
    \item We explored specific refactoring actions suggested by reviewers.
    \item We investigated two additional research questions (RQ4 and RQ5 in Section \ref{research_questions}) discussing the resolution time of smells and also the reasons why developers chose to ignore the identified smells. 
\end{enumerate}

The paper is structured as follows: related work is presented in Section \ref{sec:relatedWork}. The study design and data extraction methods are then explained in Section \ref{sec:methodology} and the results are presented in Section \ref{sec:results}; this is followed by a discussion in Section \ref{sec:discussion} before threats to the validity in Section \ref{sec:threats}; finally, conclusions and future work in Section \ref{sec:conclusion}.
\section{Related Work}
\label{sec:relatedWork}

\subsection{Studies on Code Smells}
A growing number of studies have investigated the impact of code smells on software quality, including defects \citep{Hall2014,Khomh2009}, maintenance \citep{Sjoberg2013} and program comprehension \citep{Abbes2011}. Other studies have looked at the impact of code smells on software quality using a group of developers working on a specific project \citep{Sjoberg2013,Palomba2014,Soh2016}. 

\cite{tufano2015and} mined version histories of 200 open source projects to study when code smells were introduced and the main reason behind their interaction. It was found that smells appeared in general as a result of maintenance and evolution activities. \cite{Sjoberg2013} investigated the relationship between the presence of code smells and maintenance effort through a set of control experiments. Their study did not find significant evidence that the presence of smells led to increased maintenance effort. Previous studies also include work investigating the impact of different forms of smells on software quality, such as architectural smells \citep{garcia2009,martini2018identifying}, test smells \citep{Bavota2015,Tahir2016} and spreadsheet smells \citep{Dou2014smells}.

A number of previous studies have investigated developer perception of code smells and their impact in practice.
A survey on developer perception of code smells conducted by \cite{Palomba2014b} found that developer experience and system knowledge are critical factors in the identification of code smells. \cite{Yamashita2013e} reported that developers are moderately concerned about code smells in their code. A recent study by \cite{taibi2017developers} replicated the two previous studies \citep{Yamashita2013e,Palomba2014b} and found that the majority of developers always considered smells to be harmful; however, it was found that developers perceived smells as critical in theory, but not as much in practice. \cite{tahir2020stackexchange} mined posts from Stack Exchange sites to explore how the topics of code smells and anti-patterns were discussed amongst developers. Their study found that developers widely used online forums to ask for general assessments of code smells or anti-patterns instead of asking for particular refactoring solutions.

\subsection{Code Reviews in Software Development}
Code review is an integral part in modern software development. In recent years, empirical studies on code reviews have investigated the potential code review factors that affect software quality. For example, \cite{McIntosh2014impact} investigated the impact of code review coverage and participation on software quality in the Qt, VTK, and ITK projects. The authors used the incidence rates of post-release defects as an indicator and found that poorly reviewed code (e.g., with low review coverage and participation) had a negative impact on software quality. \cite{Anderson2021predicting} investigated whether and how technical (e.g., number of times a file has been changed and types of change) and social (e.g., number of prior code changes submitted by the code owner and centrality of the code owner on the collaboration graph) metrics can be used to predict design impactful changes by analyzing more than 50k code reviews of seven real-world systems.

Some studies have focused on the impact of code review on software quality. A study by \cite{kemerer2009impact} investigated the impact of review rate on software quality. The authors found that the \textit{Personal Software Process} review rate was a significant factor affecting defect removal effectiveness, even after accounting for developer ability and other significant process variables. Several studies \citep{mcintosh2016empirical, McIntosh2014impact, Kononenko2015investigating} have investigated the impact of modern code review on software quality. Other studies have also investigated the impact of code reviews on different aspects of software quality, such as vulnerabilities \citep{bosu2014identifying}, design decisions \citep{zanaty2018empirical}, anti-patterns \citep{morales2015code} and code smells \citep{nanthaamornphong2016empirical, pascarella2020reviews}. 

Many recent studies of code review are based on pull requests (PR). \cite{wessel2020effects} conducted an empirical study on the effects of adopting bots to support the code review process on pull requests. They found that the adoption of code review bots increased the monthly number of merged pull requests with less communication between maintainers and contributors and lead projects to reject fewer pull requests. \cite{coelho2021refactoring} investigated technical aspects characterizing refactoring-inducing PRs based on data mined from GitHub and refactorings detected by RefactoringMiner. They found that PRs that induced refactoring edits have different characteristics from those that do not. Besides, their qualitative analysis indicates that at least one refactoring edit was induced by code review in more than half of refactoring-inducing PRs they studied. \cite{Cassee2020silent} present an exploratory empirical study investigating the effects of Continuous Integration (CI) on open source code reviews. They found that the number of comments per code review decreases after the adoption of CI, while the number of changes made during a code review remains constant.

\cite{panichella2020empirical} investigated the approaches and tools that are needed to facilitate code review activities from a developer point of view. They found that developers performed additional activities or tasks (e.g., the need to fix licensing and security issues) during code review with the availability of new emerging development technologies and practices, thus additional types of feedback and novel approaches and tools (e.g., for automatically detecting and fixing documentation issues) are wanted by developers.
However, code review is basically a human task involving technical, personal and social aspects. \cite{chouchen2021Anti-pattern} present the concept of Modern Code Review Anti-patterns (MCRA) and identify five common MCR anti-patterns. They conducted a study by analyzing 100 reviews randomly selected from the OpenStack project. Their preliminary results show that these anti-patterns are indeed prevalent in MCR, affecting 67\% of code reviews. 

\cite{nanthaamornphong2016empirical} examined review comments from code reviewers and described the need for an empirical analysis of the relationship between code smells and peer code review. Their preliminary analysis of review comments from OpenStack and WikiMedia projects indicated that code review processes identified a number of code smells. However, the study only provided preliminary results and did not investigate the causes of, or resolution strategies for, these smells. A more recent study by \cite{pascarella2020reviews} found that code reviews helped in reducing the severity of code smells in source code, but this was mainly a side effect to other changes unrelated to the smells themselves. 
\section{Methodology}
\label{sec:methodology}

We detail the methodology followed in this study in this section. We explain the methods used to collect, analyse and report our results for the research questions answered in this study. 

\subsection{Research Questions}
\label{research_questions}
The main goal of this study is to investigate how code smells are addressed during the course of a modern code review process. 
Specifically, we analyse code reviews for the purpose of understanding the nature of code smells in code reviews from a code reviewer and developer point of view. This goal is decomposed into the following five research questions (RQs): \\

\noindent\textbf{RQ1: Which code smells are the most frequently identified by code reviewers?}

\noindent\textbf{Rationale}: This question aims to find out the frequency with which code smells are identified by code reviewers and what particular code smells are repeatedly detected/reported by reviewers. Such information can help in improving developers' awareness of these frequently identified code smells and also help tool designers to focus on smells of more interest to developers.\\ 

\noindent\textbf{RQ2: What are the common causes for code smells that are identified during code reviews?}

\noindent\textbf{Rationale}: This question investigates the main reasons behind the identified smells as explained by the reviewers or developers. Previous research has shown that context is important in identifying code smells \citep{tahir2020stackexchange,sae2018context}.
When conducting a review, reviewers can express and explain why they think the code under review may contain a smell. Developers can also reply to reviewers and explain what they think of the smell(s), and, if they agree with the reviewers' assessment, how they introduce the smell(s). Understanding the common causes of code smells identified manually by reviewers will shed some light on the effectiveness of manual detection of smells and help developers better understand the nature of identified smells and context in which those smells are being labelled.\\

\noindent\textbf{RQ3: How do reviewers and developers treat the identified code smells?}

\noindent\textbf{Rationale}: This question investigates the actions suggested by reviewers and those taken by developers on the identified smells. When a smell is identified, reviewers can provide suggestions to resolve the smell and developers can then decide on whether to fix or ignore the code with the smell. In addition to this, we also investigate the concrete refactoring actions (e.g., move/extract method) suggested by reviewers. 
This question is further divided into three sub-questions (from the perspective of the reviewer, developer and the relationship between their actions): 

\textbf{RQ3.1: What actions do reviewers \textit{suggest} to deal with the identified smells?}

\textbf{RQ3.2: What actions do developers \textit{take} to resolve the identified smells?}

\textbf{RQ3.3: What is the \textit{relationship} between the actions suggested by reviewers and those taken by developers?} \\

\noindent\textbf{RQ4: How long does it take to resolve code smells by developers after they have been identified by reviewers?}

\noindent\textbf{Rationale}: With this question, we want to investigate the influence of different code smell categories on the fix time. In addition, combined with the results of RQ3.1, we also want to know the influence of reviewers' suggestions on the fix time of code smells. This can help in understanding the nature of each of those smells, and how difficult (using time as an indicator of difficulty) it can be to implement such fixes.

\noindent\textbf{RQ5: What are the common causes for not resolving code smells that have been identified in code?}

\noindent\textbf{Rationale}: In the case where code smells are not resolved, technical debt is introduced by developers. Consequently, with this question, we want to know what makes the developers choose to ignore the smells. Understanding this can further help to remove smells and pay back technical debt.

\subsection{Research Setting}
\label{sec:research_setting}
We conducted our study using the projects from two large and active open-source communities: OpenStack and Qt. OpenStack is a set of software tools for building and managing cloud computing platforms. It is considered one of the largest open source communities. OpenStack projects contain around 13 million lines of code, contributed to by around 12 thousand developers\footnote{As of March 2022: \url{https://www.openhub.net/p/openstack}}. Qt is an open source cross-platform application and UI framework developed by the Digia operation. Contributions form different large communities are also welcomed by Qt.

We deemed these two communities to be appropriate for our study, since they are large in size and both have long investment in their code review process. We then selected two of the most active projects (based on the number of patch submissions \citep{hirao2020code}) in each of those communities as our subject projects: Nova (a fabric controller) and Neutron (a network connectivity platform) from the OpenStack Community and Qt Base (core UI functionality) and Qt Creator (Qt IDE) from the Qt community. 

The two OpenStack projects (Nova and Neutron) are mainly written in Python, while the Qt projects (Qt Base and Qt Creator) are mostly written in C++. All these projects use Gerrit\footnote{\url{https://www.gerritcodereview.com}}, a web-based modern code review platform built on top of Git. The Gerrit review workflow is explained next.

Gerrit is designed for modern code review and provides a detailed code review workflow. First, a developer (author) makes a change to the code and submits the code (patch) to the Gerrit server so that it can be reviewed. Then, verification bots check the code using static analysers and run automated tests. A reviewer (usually other developers that have not been involved in writing the code under review) will then conduct a formal review of the code and provide comments. The original author can reply to the reviewer's comments and action the required changes by producing a new revision of the patch. This process is repeated until the change is merged to the code base or finally abandoned.

\subsection{Mining Code Review Repositories}
\label{mining_code_reivew_repositories}

\begin{figure*}[h]
    \centering
    \includegraphics[width=1.0\linewidth]{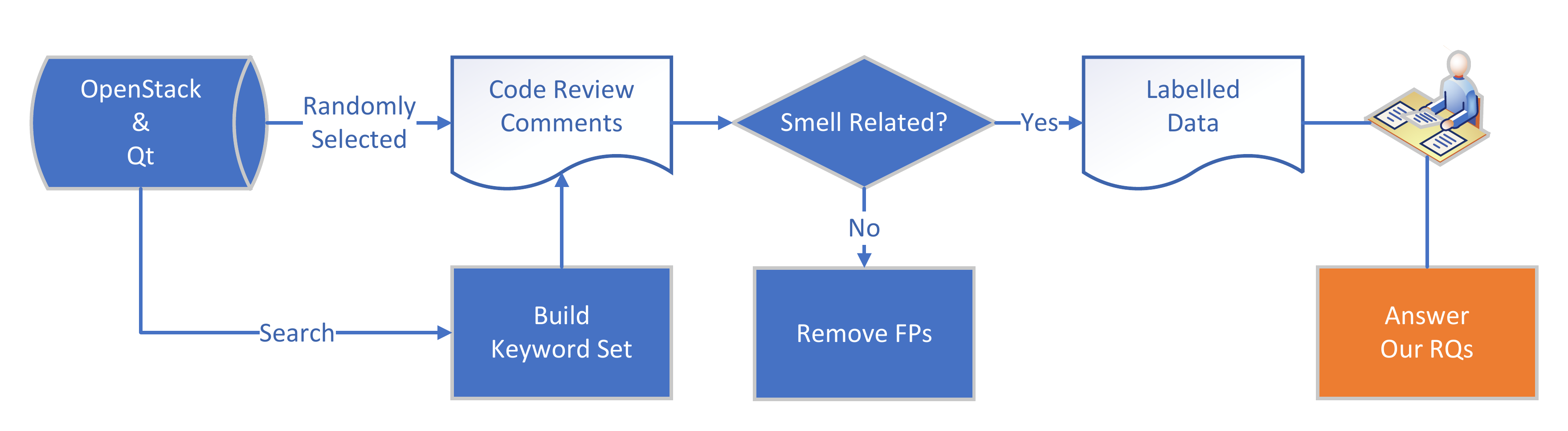}
    \caption{An overview of our data mining and analyzing process}
    \label{fig:data_mining_and_extraction}
\end{figure*}

Fig. \ref{fig:data_mining_and_extraction} outlines our data extraction and mining process. 
We mined the code review data via the RESTful API provided by Gerrit, which returns the results in a JSON format. We used a Python script to automatically mine the review data in the studied period and store the data in a local database. Details of the four projects are shown in Table \ref{tab:subject_projects}. 

\begin{table}[h]
\centering
\caption{An overview of the subject projects}
\label{tab:subject_projects}
\begin{tabular}{l|c|c|c}
\toprule
\textbf{Project} & \textbf{Review Period} & \textbf{\#Code Changes} & \textbf{\#Comments} \\ \midrule
\textbf{Nova}       & \multirow{4}{*}{Jan 2014 - Dec 2018}    & 22,762    & 156,882   \\
\textbf{Neutron}    &         & 15,256      & 152,429             \\ 
\textbf{Qt Base}    &         & 27,340      & 104,272             \\ 
\textbf{Qt Creator} &         & 28,229      & 79,087              \\ \midrule
\textbf{Total}      &         & 93,587      & 492,670             \\ \bottomrule
\end{tabular}
\end{table}

In total, we mined 93,587 code changes and 492,670 code review comments between Jan. 2014 and Dec. 2018 from the four projects. Upon checking the data, we found that there are comments published by bots in Qt Base and Qt Creator projects. Since our goal is to investigate manual detection of code smells in code review, we decided to remove all bots generated comments (51,837 comments in total). By doing so, we ended up with a total of 93,587 code changes and 440,833 review comments (309,311 from the OpenStack community and 131,522 from the Qt community) for further analysis.

\subsection{Building the Keyword Set}
\label{build_keyword_set}
To locate code review comments that include code smell discussions, we used several variations of terms referring to code smells or anti-patterns, including ``code smell'', ``bad smell'', ``bad pattern'', ``anti-pattern'' and ``technical debt''. In addition, considering that reviewers may point out a specific code smell by its name (e.g., dead code) rather than using general terms, we also included a list of specific code smell terms obtained from \cite{Tahir2018EASE}, that extracted these smell terms from several relevant studies on this topic, including the first work on code smells by \cite{martin1999refactoring} and the systematic review by \cite{zhang2011code}.
The list of these specific smell terms used in our study are shown in Table \ref{tab:smell_terms}.

\begin{table*}[h]
\centering
\caption{Specific code smell terms included in our data mining process}
\resizebox{\columnwidth}{!}{
\begin{tabular}{llll}
\toprule
\multicolumn{4}{c}{\textbf{Specific Code Smell Terms}}                                                                           \\ \toprule
Accidental Complexity               & Anti Singleton                    & Bad Naming                & Blob Class                \\
Circular Dependency                 & Coding by Exception               & Complex Class             & Complex Conditionals      \\
Data Class                          & Data Clumps                       & Dead Code                 & Divergent Change          \\
Duplicated Code                     & Error Hiding                      & Feature Envy              & Functional Decomposition  \\
God Class                           & God Method                        & Inappropriate Intimacy    & Incomplete Library Class  \\
ISP Violation                       & Large Class                       & Lazy Class                & Long Method               \\
Long Parameter List                 & Message Chain                     & Middle Man                & Misplaced Class           \\
Parallel Inheritance Hierarchies    & Primitive Obsession               & Refused Bequest           & Shotgun Surgery           \\
Similar Subclasses                  & Softcode                          & Spaghetti Code            & Speculative Generality    \\
Suboptimal Information Hiding       & Swiss Army Knife                  & Temporary Field           & Use Deprecated Components \\ \bottomrule
\end{tabular}
}
\label{tab:smell_terms}
\end{table*}

Since the effectiveness of the keyword-based mining approach relies on the set of keywords that are used in the search, we followed the systematic approach used by \cite{bosu2014identifying} to identify the keywords included in our search. This includes the following steps\footnote{implemented using the NLTK package: \url{https://www.nltk.org}}:

\begin{enumerate}[itemindent=1.5em]
    \item Build an initial keyword set (as described above).
    \item Build a corpus by searching for review comments that contain at least one keyword of our initial keyword set (e.g., ``dead'' or ``duplicated'') in the code review data we collected in Section \ref{mining_code_reivew_repositories}.
    \item Process the identified review comments which contain at least one keyword of our initial keyword set and then apply the identifier splitting rules (i.e., ``isDone'' becomes ``is Done'' or ``is\_done" becomes ``is done").
    \item Create a list of tokens for each document in the corpus.
    \item Clean the corpus by removing stopwords, punctuation and numbers and then convert all the words to lowercase.
    \item Apply the Porter stemming algorithm \citep{Porter2001Snowball} to obtain the stem of each token.
    \item Create a Document-Term matrix \citep{tan2016introduction} from the corpus.
    \item Find the additional words that co-occurred frequently with each of our initial keywords (co-occurrence probability of 0.05 in the same document).
\end{enumerate}

After performing these eight steps, we found that no additional keywords co-occurred with each of our initial keywords, based on the co-occurrence probability of 0.05 in the same document. Therefore, we believe that our initial keyword set is sufficient to support the keyword-based mining method. The initial set of keywords (which is the same as the final set of keywords) is shown in Table \ref{tab:initial_set_of_keywords}.

\begin{table}[h]
\centering
\caption{The initial set of keywords included in our data mining process}
\label{tab:initial_set_of_keywords}
\begin{tabular}{ll}
\toprule
\textbf{Code Smell Term}            & \textbf{Keywords} \\ \midrule
Code Smell                          & smell, smelly \\
Bad Smell                           & bad, smell, smelly \\
Anti-Pattern                        & anti, pattern, bad \\
Bad Pattern                         & bad, pattern \\
Technical Debt                      & technical, debt \\
Accidental Complexity               & accidental, complexity, complex \\
Anti Singleton                      & anti, singleton \\
Bad Naming                          & bad, naming \\
Blob Class                          & blob \\
Circular Dependency                 & circular, circularity, dependency, dependent \\
Coding by Exception                 & exception \\
Complex Class                       & complex, complexity \\
Complex Conditionals                & complex, complexity, conditional, condition \\
Data Class                          & data class \\
Data Clumps                         & clump \\
Dead Code                           & dead, death, unused, useless \\
Divergent Change                    & divergent, divergence \\
Duplicated Code                     & duplicated, duplicate, duplication, clone \\
Error Hiding                        & hiding, hide \\
Feature Envy                        & envy \\
Functional Decomposition            & decompose, decomposition \\
God Class                           & god, brain \\
God Method                          & god, brain \\
Inappropriate Intimacy              & inappropriate, intimacy \\
Incomplete Library Class            & incomplete, library \\
ISP Violation                       & ISP, violate, violation \\
Large Class                         & large, big \\
Lazy Class                          & lazy \\
Long Method                         & long \\
Long Parameter List                 & long, parameter list \\
Message Chain                       & chain \\
Middle Man                          & middle \\
Misplaced Class                     & misplace, misplaced \\
Parallel Inheritance Hierarchies    & parallel, inheritance \\
Primitive Obsession                 & obsession \\
Refused Bequest                     & refuse, refused, bequest \\
Shotgun Surgery                     & shotgun, surgery \\
Similar Subclasses                  & similar, subclass \\
Softcode                            & softcode \\
Spaghetti Code                      & spaghetti \\
Speculative Generality              & speculative, generality \\
Suboptimal Information Hiding       & suboptimal, hiding, hide \\
Swiss Army Knife                    & swiss, army, knife \\
Temporary Field                     & temporary, temporal \\
Use Deprecated Components           & deprecated, deprecate, component \\
\bottomrule
\end{tabular}
\end{table}

\subsection{Identifying Smell-related Reviews in Keywords-searched Review Comments}
\label{keywords_searched_review_comments}

We identified smell-related code reviews in four steps, as follows:

In \textbf{step one}, we developed a script to search for review comments that contained at least one of the keywords identified in Section \ref{build_keyword_set}. The search returned a total of 23,292 review comments from the four projects.

In \textbf{step two}, two of the authors independently and manually checked the review comments obtained in \textbf{step one} without considering any other information to exclude comments that were \emph{clearly} unrelated to code smells. When both coders decided a review comment was \emph{clearly} not related to code smells, we excluded it from any future analysis. The Cohen's Kappa coefficient value \citep{jacob1960coefficient} is 0.83, which indicates a near perfect agreement between the two coders. The number of votes is shown in Table \ref{tab:number_of_votes}. As a result of this step, the number of remaining review comments became 4,761.

To illustrate this process, consider the following two review comments that contain the keyword ``dead''. In the first example, the reviewer commented that: ``\textit{why not to put the port on dead vlan first?}''\footnote{\url{http://alturl.com/gqn7u}}. Although this comment contains the keyword ``dead'', both coders agreed that it was unrelated to code smells and the comment was therefore excluded. In the second example, the reviewer commented: ``\textit{remove dead code}''\footnote{\url{http://alturl.com/2kcko}}, which was regarded  as related to \textit{dead code} by the two coders and was included in the analysis.

In \textbf{step three}, the same two coders worked together to manually analyze the remaining (4,761) review comments using the related information of each review comment, including the code review discussions and associated source code to determine whether the code reviewers identified any smells in the review comments. We considered a comment to be related to code smells only when both coders agreed. The agreement between the two coders was calculated using the Cohen's Kappa coefficient \citep{jacob1960coefficient}, which is 0.84 (almost perfect agreement). The number of votes for TT (True-True), TF (True-False), FT (False-True), FF (False-False) from the two coders is shown in Table \ref{tab:number_of_votes}. When the coders were unsure or disagreed about the outcomes, a third author was then involved in the discussion until a consensus was reached. This resulted in a reduction in the number of review comments to 1,592.

\begin{table}[h]
\centering
\caption{The number of votes for TT, TF, FT, FF from the two coders in step two and three}
\label{tab:number_of_votes}
\begin{tabular}{c|cc|c|cc}
\toprule
\multicolumn{3}{c}{\textbf{Step two}}                       & \multicolumn{3}{c}{\textbf{Step three}} \\ \midrule
\diagbox{\textbf{Coder 2}}{\textbf{Coder 1}}  & \textbf{Maybe}    & \textbf{No}   & \diagbox{\textbf{Coder 2}}{\textbf{Coder 1}}    & \textbf{Yes}      & \textbf{No}   \\ \midrule
\textbf{Maybe}              & 3576              & 482           & \textbf{Yes}                  & 1435              & 151           \\
\textbf{No}                 & 703               & 18531         & \textbf{No}                   & 187               & 2988          \\ \bottomrule
\end{tabular}
\end{table}

\begin{figure}[htb]
    \centering
    \captionsetup{justification=centering}
    \includegraphics[width=1.0\linewidth]{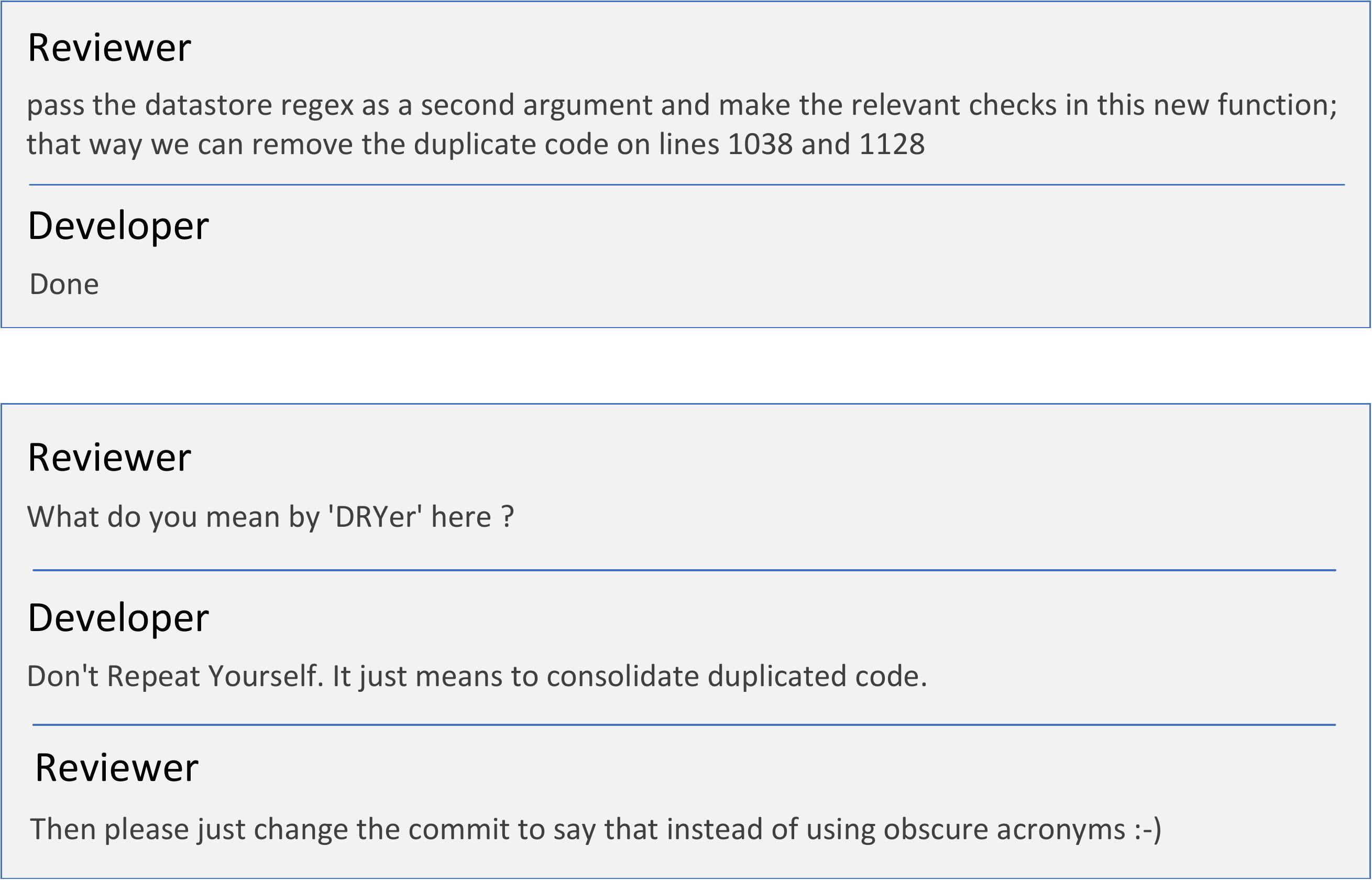}
    \caption{Review comments related to \textit{duplicated code}: the top review is smell-related, while the bottom one is not}
    \label{fig:smell_examples}
\end{figure}

To better explain our selection process, consider the two examples in Fig. \ref{fig:smell_examples}. In the top example\footnote{\url{http://alturl.com/4s775}}, the reviewer suggested adding another argument to a method to eliminate code duplication. Then the developer replied: ``\textit{Done}'', which implies an acknowledgment of the code duplication. We considered this as a clear smell-related review and the review comment was retained for further analysis. In contrast, in the bottom example\footnote{\url{http://alturl.com/786zn}}, we observed that the comment was just used to explain the meaning of the ``DRY'' principle, but did not indicate that the code contained duplication according to the context. Thus, this comment was excluded from analysis.

Finally, in \textbf{step four}, we recorded the related information of each review comment in an external text file for further analysis, which contained: 1) a URL to the code change, 2) the type of the identified code smell, 3) the discussion between reviewers and developers and 4) a URL to the source code. We ended up with a total of 1,502 smell-related reviews (we note that several review comments appearing in the same discussion were merged). An example of an extracted source file is shown below:
\\

\begin{qoutebox}{white}{}
\textbf{Code Change URL:} \url{http://alturl.com/2ne85}\\
\textbf{Code Smell:} \emph{Dead Code} \\
\textbf{Code Smell Discussions:} \\
\textbf{1) Reviewer:} ``Looks like copy-paste of above and, more importantly, dead code.'' \\
\textbf{2) Developer:} ``yes, sorry for that.''\\
\textbf{Source Code URL:} \url{http://alturl.com/yai68}
\end{qoutebox}

\subsection{Identifying Smell-related Reviews in Randomly-selected Review Comments}
\label{randomly_selected_review_comments}

Knowing that reviewers and developers may not use the same keywords as we used in Section \ref{keywords_searched_review_comments} when detecting and discussing code smells during code review, we supplemented our keyword-based mining approach by including a randomly selected set of review comments from the rest of the review comments (291,229 in the OpenStack projects and 126,312 in the Qt projects) that did not contain any of the keywords used in Section \ref{build_keyword_set}. Based on 95\% confidence level and 3\% margin of error \citep{israel1992dss}, we ended up with an additional 1,064 review comments from the OpenStack projects and 1,059 review comments from the Qt projects. We then followed the same process of manual analysis (i.e., from \textbf{step two} to \textbf{step four} as described in Section \ref{keywords_searched_review_comments}) to identify smell-related reviews in these randomly selected review comments. Finally, we identified a total of 37 smell-related reviews from the randomly selected review comments.

\begin{figure}[htb]
    \centering
    \captionsetup{justification=centering}
    \includegraphics[width=1.0\linewidth]{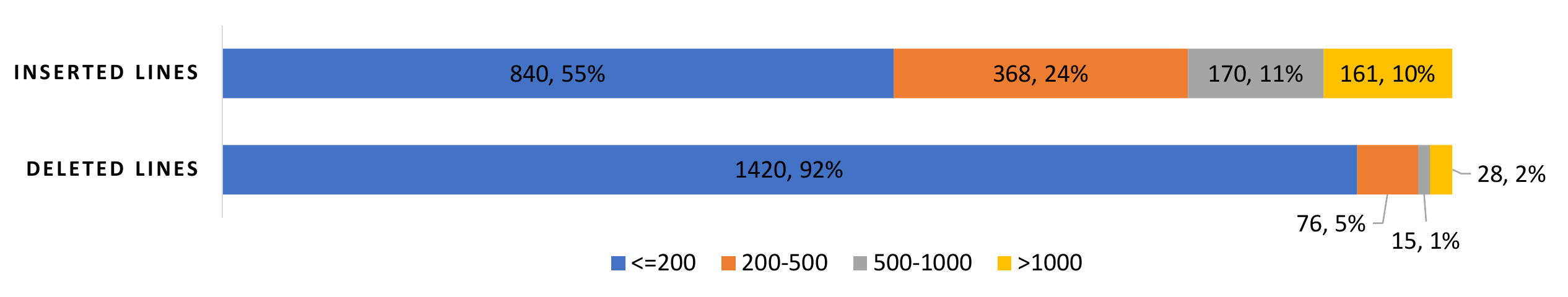}
    \caption{The size of code changes related to the identified smell-related reviews}
    \label{fig:code_changes_size}
\end{figure}

In addition to the 1,502 smell-related reviews obtained by keywords search in Section \ref{keywords_searched_review_comments}, we finally obtained a total of 1,539 smell-related reviews for further analysis. Fig. \ref{fig:code_changes_size} shows the size (in the form of inserted and deleted lines) of code changes related to the identified smell-related reviews. For inserted lines, 55\% of code changes have no more than 200 lines of newly added code. Only 21\% of the code changes have more than 500 insertions. As for deleted lines, 92\% of code changes only deleted no more than 200 lines. This suggests that the size of code changes was relatively small, generally.

\subsection{Manual Analysis and Classification}
\label{sec:manualAnalysis}

\subsubsection{RQ1: Which code smells are the most frequently identified by code reviewers?}
In Sections \ref{keywords_searched_review_comments} and \ref{randomly_selected_review_comments}, we explained how we identified and recorded the smell type noted in each review when analyzing the review comments. When a reviewer used general terms (such as ``smelly'' or ``anti-pattern'') to describe the identified smell, we classified the type in these reviews as ``general''. The others were classified as specific smells, based on the keyword included and the description provided (e.g., \textit{duplicated code}).

\subsubsection{RQ2: What are the common causes for code smells that are identified during code reviews?}
For RQ2, we adopted Thematic Analysis \citep{braun2006using} to find the causes for the identified code smells in Sections \ref{keywords_searched_review_comments} and \ref{randomly_selected_review_comments}. We used MAXQDA\footnote{\url{https://www.maxqda.com/}} - a software package for qualitative research - to code the related information of the identified code smells. Firstly, we coded the collected smell-related reviews by highlighting sections of the text related to the causes of the code smell in the review. When no cause was found, we used ``cause not provided/unknown''. Next, we looked over all the codes that we created to identify common patterns and generated themes. We then reviewed the generated themes by returning to the dataset and comparing our themes against it. Finally, we named and defined each theme. We also undertook further analysis from the perspective of smell types, to investigate the main causes leading to the introduction of a specific smell type.

This process was performed by the same two coders as in Section \ref{keywords_searched_review_comments} and Section \ref{randomly_selected_review_comments}. A third author was involved in cases of disagreement by the two coders.

\subsubsection{RQ3: How do reviewers and developers treat the identified code smells?}
For RQ3, we manually checked the code reviews obtained by following the process described in Section \ref{keywords_searched_review_comments} and Section \ref{randomly_selected_review_comments} to identify the actions suggested by reviewers and taken by developers. 

For RQ3.1, we placed the actions recommended by reviewers into three categories, as proposed in \cite{Tahir2018EASE}:

\begin{enumerate}[itemindent=1.5em]
    \item\textbf{Fix}: recommendations are made to refactor the code smells.
    \item \textbf{Capture}: detecting that there may be a code smell, but no direct refactoring recommendations are given.
    \item \textbf{Ignore}: recommendations are to ignore the identified smells. 
\end{enumerate}

When reviewers provided \textit{fix} actions, we considered this as a refactoring action. We then further investigated the concrete refactoring actions provided by reviewers based on the classification in \cite{martin1999refactoring}, which provides 7 categories with 72 specific refactorings. For example, when you have a code fragment that can be grouped together, you can use \textbf{Extract Method} (i.e., move this code to a separate new method (or function) and replace the old code with a call to the method). Another example is related to generalization. If you have two classes with similar features, you can use \textbf{Extract Superclass} to refactor your code. That is, create a superclass and move the common features to the superclass. When no specific refactoring action could be extracted from one review, we chose to exclude it from our analysis.

For RQ3.2, we investigated how developers responded to reviewers that identified code smells in their code. We conducted this analysis following a three step approach: 
We first checked the developer’s response to the reviewer in the discussion (Gerrit provides a discussion platform for both reviewers and developers). Second, we investigated the associated source code file(s) of the patch before the review was conducted and the changes in the source code made after the review.
Finally, if the developers neither responded to the reviewers nor modified the source code, we then checked the status of the corresponding code change (i.e., merged or abandoned).

We considered the identified code smells to be solved in these two cases: 1) changes were made in the source code file(s) and 2) the corresponding code change was finally abandoned (i.e., when the code change was abandoned, the code change was not be merged into the code base. In other words, the code smell no longer existed).

There were cases where developers would not fix the smell immediately and said that they would fix the identified smell in the future (i.e., in a later code change). In such a case, it is difficult to judge whether the identified smell was finally fixed. Therefore, we regarded this situation as \textit{unknown}.

For RQ3.3, based on the results of RQ3.1 and RQ3.2, we categorized the relationship between the actions recommended by reviewers and those taken by developers into the following three categories: 

\begin{enumerate}[itemindent=1.5em]
    \item A developer \textit{agreed} with the reviewer’s recommendations.
    \item A developer \textit{disagreed} with the reviewer’s recommendations, or
    \item A developer \textit{did not respond} to the reviewer’s comments.
\end{enumerate}

These three categories were then mapped into three actions: 

\begin{enumerate}[itemindent=1.5em]
    \item \textbf{Fixed the smell}: Refactoring was done and the smell was successfully removed.
    \item \textbf{Ignored the smell}: No changes were performed to the source code and the smell was finally ignored.
    \item \textbf{Unknown}: Explained above.
\end{enumerate}

In the below example, the reviewer just pointed out an instance of a \textit{dead code} smell. We categorised the suggestion of the reviewer as ``Capture''. Subsequently, the developer replied ``Done'' to the reviewer, which meant that the developer had resolved the smell. We thought that, in such a case, the developer had \textit{agreed} with the reviewer's recommendation and fixed the smell.

\begin{qoutebox}{white}{}
\textbf{Link:} \url{http://alturl.com/7u8yq}\\
\textbf{Reviewer:} ``This is dead code since you're overwriting it next.''\\
\textbf{Developer:} ``Done''
\end{qoutebox}

\subsubsection{RQ4: How long does it take to resolve code smells by developers after they have been identified by reviewers?}
\label{methodology_for_rq4}
When an identified code smell was fixed by developers, we checked the time taken for the fix. We extracted two types of time information related to the identified code smells:

\begin{enumerate}[itemindent=1.5em]
    \item \textbf{Identification Time}: we regarded the time when the reviewer published the smell-related review comment as the identification time of the smell. 
    \item \textbf{Resolution Time}: the time taken until a smell is resolved/refactored. This time can be divided into two categories:
        \begin{enumerate}[itemindent=1.5em]
            \item when the smell was fixed in a later patch, we regarded \textit{the time of uploading the patch} as the resolution time. 
            \item when the smell was not fixed in a later patch, but the code change that contained the smell was later abandoned, we regarded \textit{the time of abandoning the change} as the resolution time.
        \end{enumerate}
\end{enumerate}

\begin{figure}[h]
    \centering
    \includegraphics[width=\linewidth]{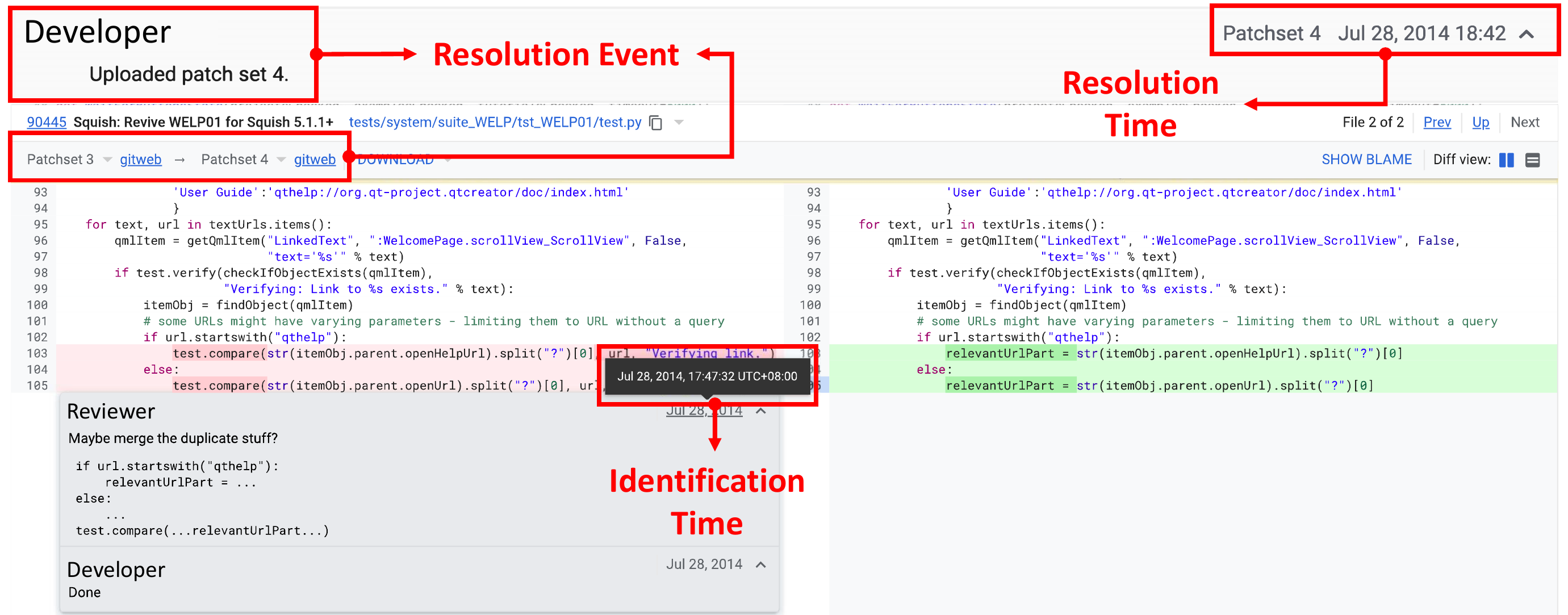}
    \caption{Example for the identification time and resolution time of a smell}
    \label{fig:example_identified_time_and_solved_time}
\end{figure}

To provide a better understanding of these two types, Fig. \ref{fig:example_identified_time_and_solved_time} shows an example from the Qt Creator project\footnote{\url{http://alturl.com/b2dkt}}. This example shows a comparison of source code between two patches: patch 3 (left) and patch 4 (right). In this example, the reviewer identified duplication (i.e., \textit{duplicated code}). We regarded the published time of this comment (i.e., Jul 28, 2014, 17:47:32 UTC+08:00) as the identification time of this smell. In patch 4, we see that the developer made a change as the reviewer suggested, implying that the smell was fixed. So we considered the upload time of patch 4 (i.e., Jul 28, 2014, 18:42:38 UTC+08:00) as the resolution time of this smell. The interval between these two time points (55 minutes and 6 seconds) is seen as the time taken to fix the code smell.

In four reviews, the time interval between the resolution time and the identification time is very long, exceeding one year. In this case, the code change where the identified smell was located took a long time to be abandoned. We believe that this case is abnormal (i.e., outliers) and including these reviews would have affected our results; consequently we excluded these four reviews.

\begin{figure}[h]
    \centering
    \includegraphics[width=\linewidth]{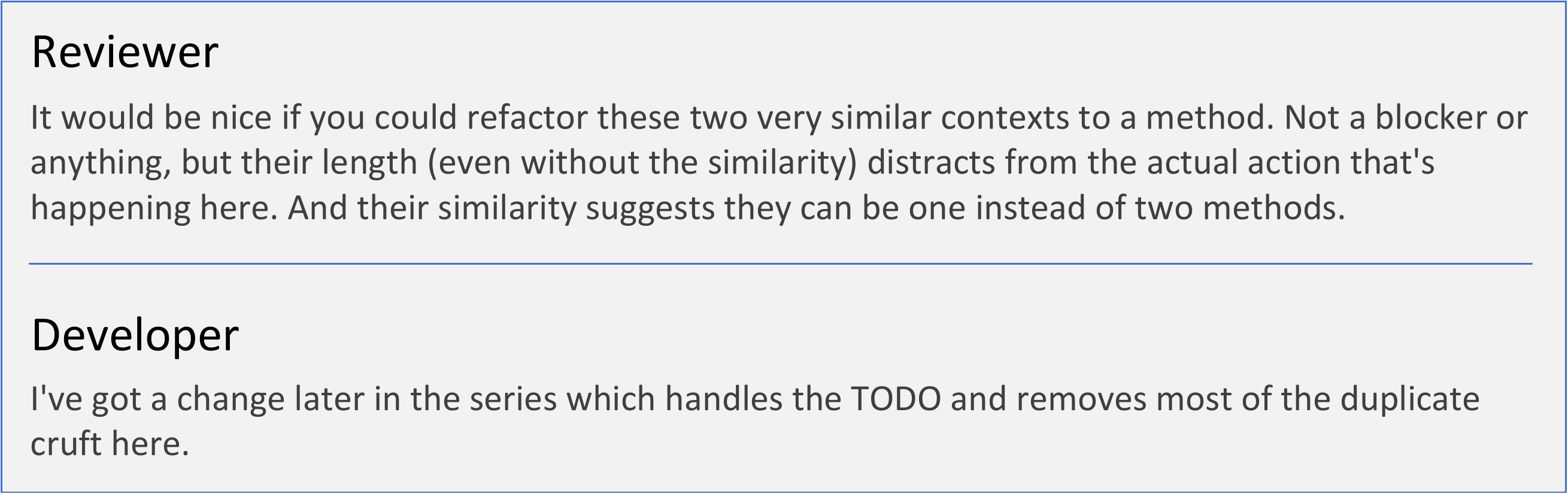}
    \caption{An example of a smell where the resolution time cannot be found}
    \label{fig:example_resolution_time_not_sure}
\end{figure}

There are also two reviews in which the resolution time of the identified smells could not be determined. Fig. \ref{fig:example_resolution_time_not_sure} shows an example from the Nova project\footnote{\url{http://alturl.com/odfjv}}. In this example, the developer replied to the reviewer that they had removed most of the duplicate code (i.e., the smell was fixed). However, we could not find out in which code change this code smell was fixed. That is, we could not get the resolution time of this smell; consequently, we excluded this review from our analysis.

There are also two reviews in which the reviewer added the smell-related comment after the code change had been abandoned. In this case, the identification time was earlier than the resolution time, which is not suitable for our analysis and consequently we excluded these two reviews too.

Finally, we decided to keep only the smell categories that had over 100 instances. Some code smell categories (i.e., \textit{long method}, \textit{circular dependency}, \textit{speculative generality}, \textit{swiss army knife} and general smell) were rarely identified and consequently these smell categories may have limited statistical significance. We decided to exclude them (35 reviews) from answering this RQ.

Moreover, to investigate the relationship between the time for fixing smells and reviewer suggestions, we divided the selected reviews into two groups: 1) reviews in which reviewers provided specific refactoring actions and 2) reviews in which reviewers just captured the smells or provided general actions. We then calculated and compared the minimum, quartile, maximum, and median time in these two groups.

\subsubsection{RQ5: What are the common causes for not resolving code smells that have been identified in code?}
For RQ5, we further investigated the reasons why developers explicitly disagreed with reviewers' assessment of smells (i.e., when developers challenge the reviewers' assessments and then decide to ignore the identified smells). We adopted Thematic Analysis \citep{braun2006using} to identify the causes of developers ignoring code smells identified by reviewers by studying the content in the reply section of the review. We followed the same process in this RQ as the one used for answering RQ2. For these cases, we then further checked the final status of the code changes to investigate what happened after developers disagreed with reviewers and decided not to fix the identified code smells.

Note that all of the manual analysis and classification (i.e., identifying smell-related code reviews and their classifications in various aspects) was conducted by at least two authors. A third author was involved in case of disagreement. In total, the manual analysis process took around 65 days (full-time) work of the coders. To facilitate replication, we provide the full data (coded) together with scripts used to collect the dataset in our replication package \citep{anonymous_replication_package}.
\section{Results}
\label{sec:results}
In this section, we present the results of our five research questions (RQs). We also provided a replication package online which is complementary for understanding the results and replicating this study \citep{anonymous_replication_package}.

\subsection{RQ1: Which code smells are the most frequently identified by code reviewers?}
\label{results_of_RQ1}

\begin{figure}[h]
    \centering
    \includegraphics[width=\linewidth]{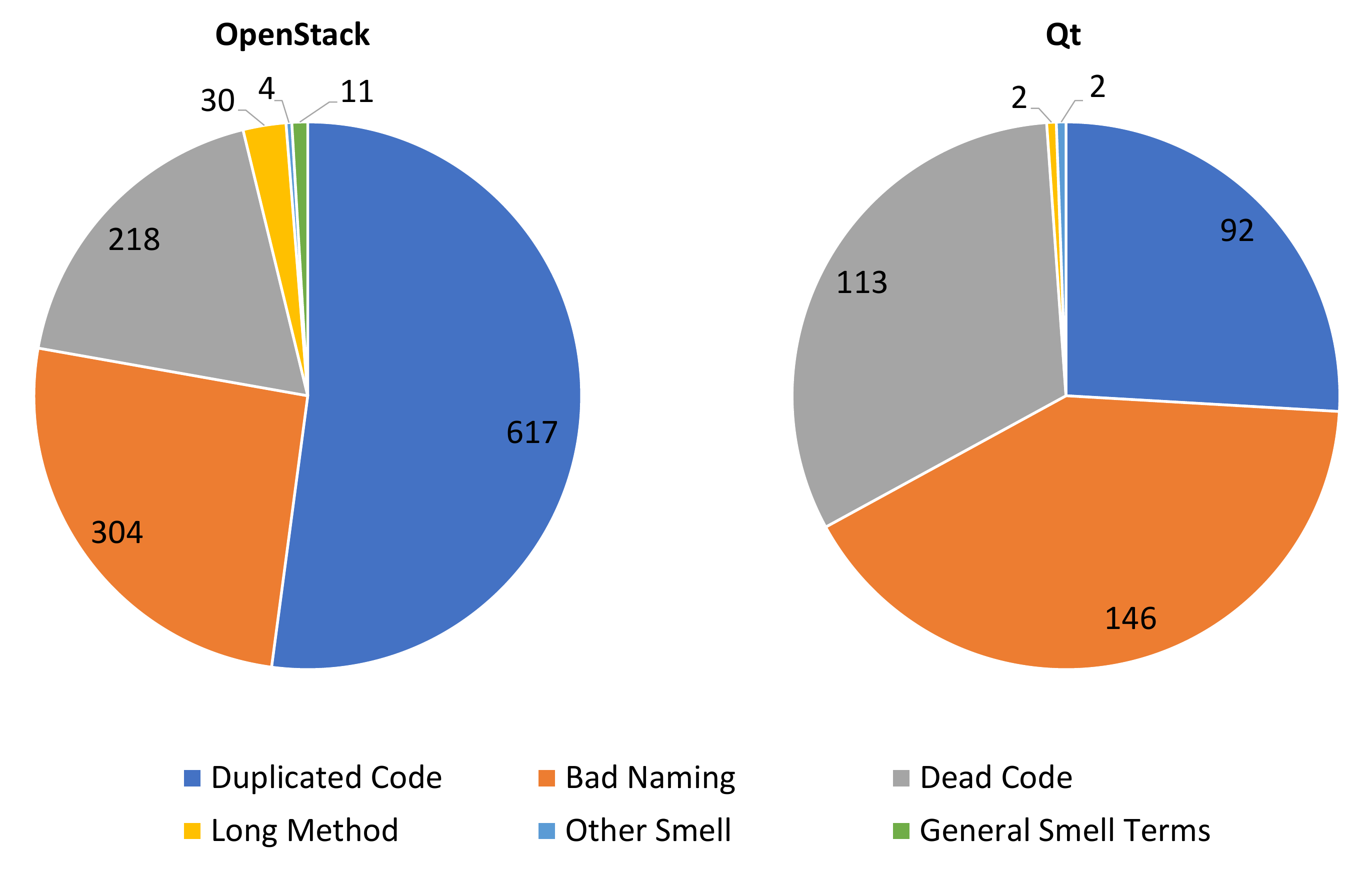}
    \caption{Distribution of smell-related reviews from the OpenStack and Qt projects}
    \label{fig:smell_distribution}
\end{figure}

We show the distribution of code smells identified in the code reviews from the OpenStack projects in Figure \ref{fig:smell_distribution}. In general, we identified 1,184 smell-related reviews in OpenStack. Of all the code smells we identified, \emph{duplicated code} is the most frequently identified smells, with exactly 617 (52\%) instances. The smells of \emph{bad naming} and \emph{dead code} are also frequently identified, as they are discussed in 304 (26\%) and 218 (18\%) code reviews, respectively. There are 30 (2\%) code reviews which identified \emph{long method}, while other smells such as \emph{circular dependency} and \emph{swiss army knife} are discussed in only 4 code reviews. The rest of code reviews (11, 1\%) use general terms (e.g., code smell) to describe the identified smells, which are called \emph{general smell terms} in this work. Note that the results of distribution of smell-related reviews from the OpenStack projects are slightly different from the results in our previous work \citep{xiaofeng2021Understanding} because we found that some code smells were mentioned by developers and were not identified through code review, which were consequently removed from our analysis.

The distribution of code smells identified in the Qt projects is shown in Fig. \ref{fig:smell_distribution}. In general, we identified 355 smell-related reviews in Qt. Unlike OpenStack, \emph{bad naming} is the most frequently identified smell, appearing in 146 (41\%) reviews. \emph{Dead code} and \emph{duplicated code} follow closely, identified in 113 (32\%) and 92 (26\%) reviews, respectively. \emph{Long method}, \emph{circular dependency}, and \emph{speculative generality} are discussed in only 2, 1 and 1 reviews, respectively. Another finding is that no \emph{general smell term} is identified in the code reviews of Qt projects. \\

\noindent\resizebox{\columnwidth}{!}{\fbox{
	\parbox{\columnwidth}{
		\textbf{RQ1 Summary:} According to the percentage of smell-related review comments (less than 1\% of total review comments), only a small number of code smells were extracted in the code review data we analysed. Of the identified smells, \textbf{\emph{duplicated code}}, \textbf{\emph{bad naming}}, and \textbf{\emph{dead code}} are the most frequently identified smells in code reviews.
	}
}}

\subsection{RQ2: What are the common causes for code smells that are identified during code reviews?}
\label{results_of_RQ2}

For RQ2, we used Thematic Analysis to identify the common causes for the identified code smells as noted by code reviewers or developers. We then identified four key causes:

\begin{itemize}
    \item \textbf{Violation of coding conventions}: certain violations of coding conventions (e.g., naming convention) are the cause for the smells. (Example: ``\emph{moreThanOneIp (CamelCase) is not our naming convention}'' \footnote{\url{http://alturl.com/azijc}}).
    \item \textbf{Lack of familiarity with existing code}: developers introduced the smells due to the unfamiliarity with the functionality or structure of the existing code. (Example: ``\emph{this useless line because \texttt{None} will be returned by default}'' \footnote{\url{http://alturl.com/h2bpc}}).
    \item \textbf{Unintentional mistakes of developers}: developers forgot to fix the smells or introduced the smells by mistake. (Example: ``\emph{You can see I renamed all of the other test methods and forgot about this one}'' \footnote{\url{http://alturl.com/cisgv}}).
    \item \textbf{Design choices}: the smells were considered to be caused by the design choice of developers. (Example: ``\emph{...If that's the case something is smelly (too coupled)...}'' \footnote{\url{http://alturl.com/y2ndw}}).
\end{itemize}

\begin{figure}[htb]
    \centering
    \includegraphics[width=\linewidth]{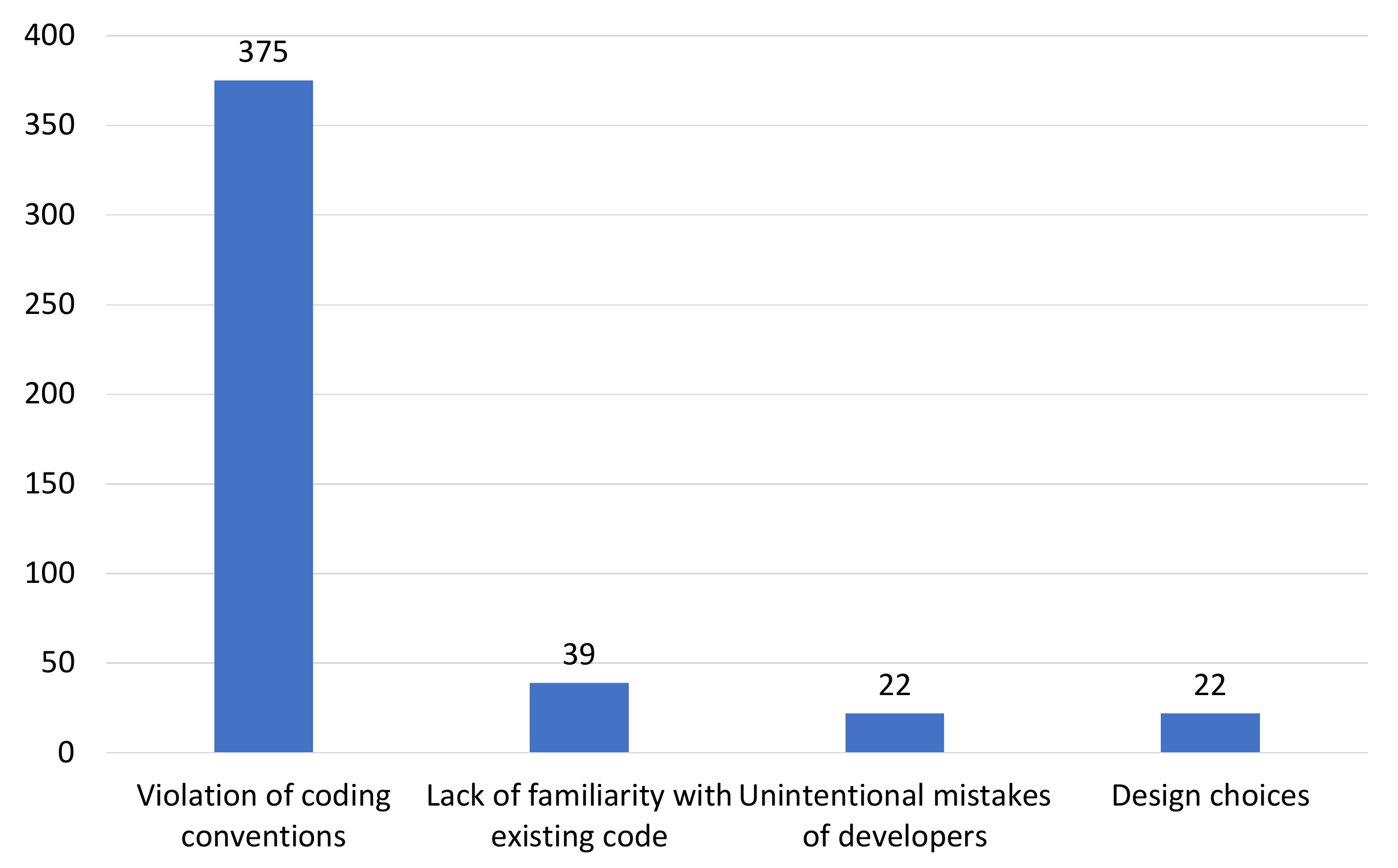}
    \caption{Causes for the identified smells (we note that there are 1,071 reviews where the reason was not provided)}
    \label{fig:reasons_for_smells}
\end{figure}

We firstly found that the majority of reviews (1,081, 70\%) did not provide any explanation for the identified smells - in most cases, the reviewer(s) simply pointed out the problems, but did not provide any further reasoning for their judgements. The detailed result is shown in Fig. \ref{fig:reasons_for_smells}.

Of the remaining 458 reviews in which the causes of the smells are provided, 375 (82\%) of the reviews indicate that \textbf{violation of coding conventions} is the main reason for the smell. For example, a reviewer suggested that the developer should adhere to the naming standard of `test\_[method under test]\_[detail of what is being tested]' which indicated a \textit{bad naming} smell, as shown below:

\begin{qoutebox}{white}{}
\textbf{Link:} \url{http://alturl.com/zw5e6} \\
\textbf{Reviewer:} ``Please adhere to the naming standard of `test\_[method under test]\_[detail of what is being tested]' to ensure that future maintainers will have an easier time associating tests and the methods they target.''
\end{qoutebox}

In addition, 39 (8\%) of the reviews indicate that the smells are caused by developers' \textbf{lack of familiarity with existing code}. An example of such a case is shown below. In this case, the reviewer pointed out that the exception did not raise, so the exception handling code became a \textit{dead code} smell that should be removed. This could imply that the developer was not aware that the specific exception was not raised.

\begin{qoutebox}{white}{}
\textbf{Link:} \url{http://alturl.com/ccjy3} \\
\textbf{Reviewer:} ``on block\_device.BlockDeviceDict.from\_api(), exception.InvalidBDMVolumeNotBootable does not raise. so it is necessary to remove the exception here.''
\end{qoutebox}

Twenty-two reviews (5\%) attribute \textbf{unintentional mistakes of developers} (such as copy and paste) to be the cause of the smells, similar to the example shown below:

\begin{qoutebox}{white}{}
\textbf{Link:} \url{http://alturl.com/zwz2x}\\
\textbf{Reviewer:} ``I think you forgot to remove this.''\\
\textbf{Developer:} ``Darn, yes bad copy / paste.  Will fix it.''
\end{qoutebox}

Twenty-two reviews (5\%) indicate that \textbf{design choices} was the cause of the identified smells. This means that the developers made a poor design choice which introduced the smell. Below is an example in which the reviewer pointed out that the code may indicate that the developer improperly decomposed some test methods.

\begin{qoutebox}{white}{}
\textbf{Link:} \url{http://alturl.com/9ctor}\\
\textbf{Reviewer:} ``The fact that this is now a one liner feels like code smell. I'm not sure, but it may indicate improper decomposition of some of these test methods.''\\
\textbf{Developer:} ``I think I see where you're coming from with this. I'm going to have another look.''
\end{qoutebox}

More specifically, from the perspective of code smell types, the distribution of causes for different smell types are shown in Table \ref{tab:distribution_of_causes}.

\begin{table}[h]
\centering
\caption{The distribution of causes for different smell types}
\label{tab:distribution_of_causes}
\begin{tabular}{l l c c}
    \toprule
    \textbf{Code Smell}     & \textbf{Cause}    & \textbf{Count}    & \textbf{\%} \\
    
    \midrule
    \multirow{3}{*}{Duplicated Code}    & cause not provided                        & 683           & 96.3\%    \\
                ~                       & lack of familiarity with existing code    & 19            & 2.7\%     \\
                ~                       & unintentional mistakes of developers      & 7             & 1.0\%     \\
    \midrule
    
    \multirow{4}{*}{Bad Naming}         & violation of coding conventions           & 368           & 81.8\%    \\
                ~                       & cause not provided                        & 79            & 17.6\%    \\
                ~                       & unintentional mistakes of developers      & 2             & 0.4\%     \\
                ~                       & lack of familiarity with existing code    & 1             & 0.2\%     \\
    \midrule
    
    \multirow{5}{*}{Dead Code}          & cause not provided                        & 284           & 85.8\%    \\
                ~                       & lack of familiarity with existing code    & 19            & 5.7\%     \\
                ~                       & design choices                            & 15            & 4.5\%     \\
                ~                       & unintentional mistakes of developers      & 13            & 4.0\%     \\
    \midrule
    
    \multirow{2}{*}{Long Method}        & cause not provided                        & 29            & 90.6\%    \\
                ~                       & design choices                            & 3             & 9.4\%     \\
    \midrule
    
    Circular Dependency                 & cause not provided                        & 4             & 100\%     \\
    \midrule
    
    Swiss Army Knife                    & cause not provided                        & 1             & 100\%     \\
    \midrule
    
    Speculative Generality              & cause not provided                        & 1             & 100\%     \\
    \midrule
    
    \multirow{2}{*}{General Smell}     & violation of coding conventions           & 7             & 63.6\%    \\
                ~                       & design choices                            & 4             & 36.4\%    \\
    \bottomrule
\end{tabular}
\end{table}

For \textit{duplicated code}, the majority of reviews (683 out of 709, 96.3\%) did not provide any cause for the identified smells. When causes are provided, \textbf{lack of familiarity with existing code} is the main cause for this smell type, accounting for 2.7\%. The situation of \textit{dead code} is similar to that of \textit{duplicated code}. In 85.8\% of the reviews, no further explanation was provided. When the cause was provided, \textbf{lack of familiarity with existing code}, \textbf{design choices} and \textbf{unintentional mistakes of developers} account for almost the same proportion, 5.7\%, 4.5\%, and 4.0\%, respectively. On the other hand, \textit{bad naming} was different to \textit{duplicated code} and \textit{dead code} smells. The reviewers usually provided an explanation for why they think a bad naming smell existed and \textbf{violation of coding conventions} is the main noted cause for \textit{bad naming}. In only 17.6\% of reviews was the cause of the \textit{bad naming} smell not provided. For the other 49 smells, 71\% of the reviews did not provide the cause of the identified smell. Both \textbf{design choices} and \textbf{violation of coding conventions} are mentioned in 7 reviews. \\

\noindent\resizebox{\columnwidth}{!}{\fbox{
	\parbox{\columnwidth}{
	\textbf{RQ2 Summary:} In general, over half of the reviews did not provide an explanation of the causes of the smells. In terms of the causes, \textbf{violation of coding conventions} is the main cause for the smell as noted by reviewers and developers. Specifically, \textbf{violation of coding conventions} is the main cause of the \textit{bad naming} smell. For the other smells, \textbf{lack of familiarity with existing code} and \textbf{unintentional mistakes of developers} are the main noted causes of smells.
	}
}}

\subsection{RQ3: How do reviewers and developers treat the identified code smells?}
\label{results_of_RQ3}

\subsubsection{RQ3.1: What actions do reviewers suggest to deal with the identified smells?}

\begin{table}[h]
\caption{Actions recommended by reviewers to resolve smells in the code}
\centering
\label{tab:actions_reviewers}
\resizebox{0.9\columnwidth}{!}{
    \begin{tabular}{l|c}
    \toprule
    \textbf{Reviewer's recommendation}                          & \textbf{Count} \\ \midrule
    Fix (without recommending any specific implementation)      & 718            \\
    Fix (provided specific implementation)                      & 405            \\
    Capture (just noted the smell)                              & 366            \\
    Ignore (no side effects)                                    & 50             \\ \bottomrule
    \end{tabular}}
\end{table}

The results of this RQ are shown in Table \ref{tab:actions_reviewers}.
In the majority of reviews (1,123, 73\%), reviewers recommended \emph{fix} resolving the identified code smells. These fixes included either general directions (such as the name of a refactoring technique to be used) or specific actions (pointing to specific changes to the code base that could remove the smell). 405 (36\%) of these fixes provided example code snippets to help developers better refactor the smells. An example review where the reviewer suggested a general \textit{fix} action is shown below.

\begin{qoutebox}{white}{}
\textbf{Link:} \url{http://alturl.com/3r3pu}\\
\textbf{Reviewer:} ``remove dead code''\\
\textbf{Developer:} ``Done''
\end{qoutebox}

Next is an example of a review that suggested a \emph{fix} recommendation with specific implementation. In this example, the reviewer suggested removing duplicate code from a test case and also provided a working example of how to apply \textbf{Extract Method} (i.e., the process of moving part of the code inside a method/function to a separate new method and replacing the existing code with a call to the newly created method) refactoring to define a new test method.

\begin{qoutebox}{white}{}
\textbf{Link:} \url{http://alturl.com/c3g69}\\
\textbf{Reviewer:} ``I think you can do function that remove duplicated code, something like that following...''
\begin{lstlisting}[language=Python,breaklines=true,basicstyle=\small]
def _compare(self, exp_real):
    for exp, real in exp_real:
        self.assertEqual(exp['count'], real.count)
        self.assertEqual(exp['alias_name'], real.alias_name)
        self.assertEqual(exp['spec'], real.spec)
\end{lstlisting}
\end{qoutebox}

366 reviews (24\%) fall under the \textit{capture} category. In those reviews, reviewers just pointed out the presence of the smells, but did not provide any refactoring suggestions. 
In a small number of reviews (50, 3\%), reviewers suggested ignoring the code smells found in the code reviews. In such a case, reviewers indicated that they could tolerate the identified code smell or there was no need to fix the smell at that point. We further analysed the types of the identified smells when reviewers suggested an \textit{ignore} action. The detailed results are shown in Table \ref{tab:ignore_smell_types}. Of the code smells that reviewers suggested ignoring, \textit{duplicated code} makes up the majority (68\%). \textit{Bad naming} follows, accounting for 22\%. The remaining code smells (i.e., \textit{dead code}, \textit{circular dependency} and \textit{long method}) only appear in 5 (10\%) reviews.

\begin{table}[h]
    \centering
    \caption{Types of code smells that reviewers suggested \textit{ignore} action}
    \label{tab:ignore_smell_types}
    \begin{tabular}{l c c}
        \toprule
        \makebox[0.3\columnwidth][l]{\textbf{Code Smell}}         & \makebox[0.2\columnwidth][c]{\textbf{Count}}        & \makebox[0.1\columnwidth][c]{\textbf{\%}}    \\
        \midrule
        Duplicated Code             & 34                    & 68\%  \\
        Bad Naming                  & 11                    & 22\%  \\
        Dead Code                   & 3                     & 6\%   \\
        Long Method                 & 1                     & 2\%   \\
        Circular Dependency         & 1                     & 2\%   \\
        \bottomrule
    \end{tabular}
\end{table}

We then investigated the specific refactoring actions provided by reviewers in cases where they recommended a \textit{fix} action. Of the 1,123 reviews where reviewers provided \textit{fix} suggestions, there are 754 (67\%) reviews where no specific refactoring actions are provided by reviewers. For smells that are straightforward to resolve, such as \textit{dead code} and \textit{bad naming}, reviewers usually just suggested removing the smell without providing concrete refactoring actions in these reviews. In the remainder of the reviews, the distribution of the specific refactoring actions recommended by reviewers is shown in Table \ref{tab:specific_refactoring_actions}.

\begin{table}[h]
    \caption{Specific refactoring actions recommended by reviewers to resolve the identified smells}
    \label{tab:specific_refactoring_actions}
    \resizebox{\columnwidth}{!}{
        \begin{tabular}{l|l|c}
            \toprule
            \textbf{Code smell}     & \textbf{Reviewer's recommendation}      & \textbf{Count} \\ 
            \midrule
            
            \multirow{10}{*}{Duplicated Code}    & Extract Method               & 261   \\
            ~ & Consolidate Duplicate Conditional Fragments                     & 34    \\
            ~ & Extract Superclass                                              & 12    \\
            ~ & Consolidate Conditional Expression                              & 12    \\
            ~ & Parameterize Method                                             & 11    \\
            ~ & Pull Up Method                                                  & 5     \\
            ~ & Extract Variable                                                & 4     \\
            ~ & Substitute Algorithm                                            & 2     \\
            ~ & Extract Subclass                                                & 2     \\
            ~ & Add Parameter                                                   & 1     \\
            \midrule

            \multirow{2}{*}{Long Method} & Extract Method                       & 19    \\
            ~ & Consolidate Duplicate Conditional Fragments                     & 2     \\
            \midrule
            
            Swiss Army Knife    & Extract Method                                & 1     \\
            \midrule
            
            General Smell   & Extract Method                                    & 3     \\
            \bottomrule
        \end{tabular}
    }
\end{table}

From the table, we can see that reviewers usually suggested concrete refactoring actions for \textit{duplicated code}, such as \textbf{Extract Method} and \textbf{Consolidate Duplicate Conditional Fragments}. It also indicates that there are multiple types of refactoring actions to solve \textit{duplicated code}. Furthermore, \textbf{Extract Method} is the most frequently suggested refactoring action, especially for fixing \textit{duplicated code} and \textit{long method}. Below we present the detailed results of \textit{duplicated code} and \textit{long method} refactoring actions: \\

\noindent\textbf{Specific refactoring actions for fixing duplicated code}\\

In cases where reviewers named specific refactoring actions (344 reviews), \textbf{Extract Method} is the most frequently suggested (in 261 reviews, 76\%) refactoring. Below is an example of where the reviewer suggested extracting the duplicated code to a new private method.

\begin{qoutebox}{white}{}
\textbf{Link:} \url{http://alturl.com/482nn}\\
\textbf{Reviewer:} ``Instead of duplicating the function body, use a \_data() function to supply test rows such as (cipher, presence of the ephemeral key)''
\end{qoutebox}

\textbf{Consolidate Duplicate Conditional Fragments} (i.e., move the code which can be found in all branches of a conditional outside of the conditional) and \textbf{Consolidate Conditional Expression} (i.e., consolidate all conditionals that lead to the same result or action in a single expression) are recommended in 34 (10\%) and 12 (3\%) reviews, respectively to solve duplication in conditional statements. Below is an example from Qt Base project, where the reviewer provided specific code snippets by consolidating duplicate conditional fragments.

\begin{qoutebox}{white}{}
\textbf{Link:} \url{http://alturl.com/on6ts}\\
\textbf{Reviewer:} ``You can remove the duplicate rect.setSize() by re-ordering these conditions.''
\begin{lstlisting}[language=c++,breaklines=true,basicstyle=\small]
    if (rect.width() < ...) {
        if (rect.isEmpty() && (touch->device...)
           diameter = ...
        rect.setSize
    }
\end{lstlisting}
\end{qoutebox}

We note that \textbf{Extract Superclass} (i.e., create a shared superclass for the classes with common fields and methods and move all the identical fields and methods to the superclass), \textbf{Pull Up Method} (i.e., make the methods that perform similar work in subclasses identical and then move them to the relevant superclass) and \textbf{Extract Subclass} (i.e., create a subclass and use it in cases where a class has features that are used only in certain cases) were suggested in 12, 5, and 2 reviews, respectively. These three refactoring actions were used to deal with generalisation in classes (e.g., two or more classes with common fields and methods that can be grouped together and the original classes extend the newly created superclass). Below is an example from the Nova project:

\begin{qoutebox}{white}{}
\textbf{Link:} \url{http://alturl.com/z6svv}\\
\textbf{Reviewer1:} ``A lot of these methods appear to be duplicated from LibvirtISCSIVolumeDriver.  Maybe just use it as the base class for LibvirtISERVolumeDriver?''\\
\textbf{Reviewer2:} ``I agree with `Reviewer1'. This is a very good point.''\\
\textbf{Developer:} ``I'm working on that.. thanks.''
\end{qoutebox}

\textbf{Parameterize Method} (i.e., combine methods that perform similar actions that are different only in their internal values, numbers or operations by using a parameter that will pass the necessary special value) and \textbf{Add Parameter} (create a new parameter to pass the necessary data for the method which does not have enough data to perform certain actions) were suggested in 11 and 1 reviews, respectively. When multiple methods perform similar actions differing only in their internal values, numbers or operations, these refactoring actions can be used (similar to code duplication) to combine these methods by using a parameter (argument) passed as a value to the method. An example of the \textbf{Parameterize Method} refactoring in the Neutron project is shown below:

\begin{qoutebox}{white}{}
\textbf{Link:} \url{http://alturl.com/o3r38}\\
\textbf{Reviewer:} ``The same comment applies to the below. The methods below are very similar and it would be better to define a common method which takes "resource" name as an argument.''\\
\textbf{Developer:} ``Done''
\end{qoutebox}

There are four reviews in which the reviewers suggest \textbf{Extract Variable} to remove the \textit{duplicated code}. In another two reviews, the \textit{duplicated code} was caused by bad (algorithm) implementation and the reviewers suggested using \textbf{Substitute Algorithm} as a solution.\\

\noindent\textbf{Specific refactoring actions for fixing long method}\\

Of the 21 reviews where the reviewers provided specific refactoring actions, \textbf{Extract Method} was also suggested as a way of extracting the \textit{long method} (appearing in 19 reviews). Below is an example of a review where the reviewer suggested splitting the code to separate (more manageable) methods.

\begin{qoutebox}{white}{}
\textbf{Link:} \url{http://alturl.com/33dvx}\\
\textbf{Reviewer:} ``Aside: this function is crazy long and could probably benefit from some refactoring into smaller helper functions.''\\
\textbf{Developers:} ``Agreed, I'll make a note to revisit during R.''
\end{qoutebox}

There are two reviews in which the reviewers suggested \textbf{Consolidating Duplicate Conditional Fragments} to refactor a \textit{long method}.

\subsubsection{RQ3.2: What actions do developers take to resolve the identified smells?}

Table \ref{tab:fixed_smells} provides details of the number of reviews that identified code smells versus the number of fixes of the identified code smells.

\begin{table}[h]
\caption{Developers' actions to code smells identified in code reviews}
\label{tab:fixed_smells}
\resizebox{\columnwidth}{!}{
\begin{tabular}{l|c|c|c}
\toprule
\textbf{Code smell}     & \textbf{\#Reviews }       & \textbf{\#Fixed by developers}    & \textbf{\% of fixes}   \\ \midrule
Duplicated Code         & 709                       & 561                               & 79\%      \\
Bad Naming              & 450                       & 400                               & 89\%      \\
Dead Code               & 331                       & 307                               & 93\%      \\
Long Method             & 32                        & 24                                & 75\%      \\
Circular Dependency     & 4                         & 2                                 & 50\%      \\ 
Swiss Army Knife        & 1                         & 1                                 & 100\%     \\
Speculative Generality  & 1                         & 1                                 & 100\%     \\
General Smell           & 11                        & 8                                 & 73\%      \\ \hline
\textbf{Total}          & \textbf{1,539}            &\textbf{1,304}                     & \textbf{85\%}     \\ \bottomrule
\end{tabular}}
\end{table}

Of the 1,539 code smells identified in the reviews, the majority (1,304, 85\%) were refactored by developers after the review (i.e., changes were made to the patch). As per the results of RQ1, \emph{duplicated code}, \emph{bad naming}, and \emph{dead code} were the most frequently identified smells by reviewers. Subsequently, those smells were also widely resolved by developers. 561 (79\%) \emph{duplicated code}, 400 (89\%) \emph{bad naming} and 307 (93\%) \emph{dead code} smells were refactored by developers after they were identified in reviews.
The proportion of other smells being fixed are 73\% (36/49).

\begin{figure*}[ht]
    \centering
    \includegraphics[width=1\linewidth]{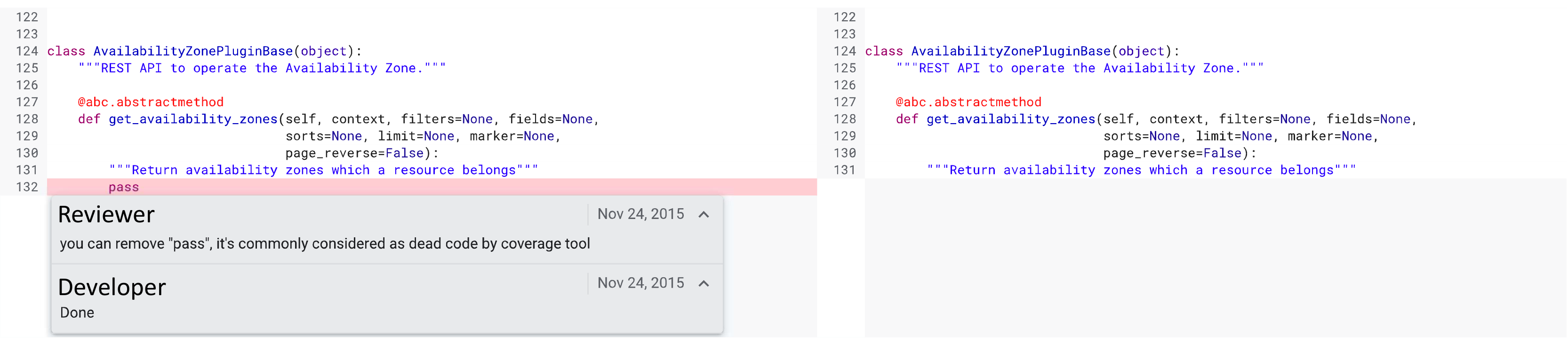}

    \subfloat[\label{fig:deadcode:before} method before review]{\hspace{.5\linewidth}}
    \subfloat[\label{fig:deadcode:after} after change made by the developer]{\hspace{.5\linewidth}}
    \caption{An example of a \emph{remove dead code} operation after review (the change is highlighted in Line 132 (a))}
    \label{fig:deadcode_example}
\end{figure*}

Below is an example of a review with a recommendation by the reviewer to remove \emph{dead code} in Line 132 of the original file (i.e., remove the \texttt{pass} statement); the developer then agreed with the reviewer's recommendation and deleted the unused code. Fig. \ref{fig:deadcode_example} shows the code before review (Fig. \ref{fig:deadcode:before}) and after the action taken by the developer (Fig. \ref{fig:deadcode:after}). 

\begin{qoutebox}{white}{}
\textbf{Link:} \url{http://alturl.com/szswu}\\
\textbf{Reviewer:} ``you can remove `pass', it's commonly considered as dead code by coverage tool''\\
\textbf{Developer:} ``Done''
\end{qoutebox}

There are 28 (2\%) reviews in which the developers indicated that they would fix the identified smells in a later change or commit. In this case, it is difficult to determine whether the identified smells are fixed or not. We categorized these cases as \textit{Unknown}. Below is an example where the developer promised to remove the duplication in another patch.

\begin{qoutebox}{white}{}
\textbf{Link:} \url{http://alturl.com/dqev8}\\
\textbf{Reviewer:} ``nit: I'm pretty sure we have this same pattern in several places in this driver code, we should create a libvirt.utils helper method for this at some point, i.e. is\_parallels(vm\_mode=None).'' \\
\textbf{Developer:} ``Ok, I'll do it in another patch''
\end{qoutebox}

The remaining 207 (13\%) reviews do not lead to any changes in the code, indicating that developers may have chosen to ignore such recommendations. This could be a case where the developers thought that those smells were not as harmful as suggested by the reviewers, or that there were other issues requiring more urgent attention, resulting in those smells being counted as technical debt in the code \citep{li2015systematic}.

\subsubsection{RQ3.3: What is the relationship between the actions suggested by reviewers and those taken by developers?}

\begin{figure}[htb]
    \centering
    \captionsetup{justification=centering}
    \includegraphics[width=1\linewidth]{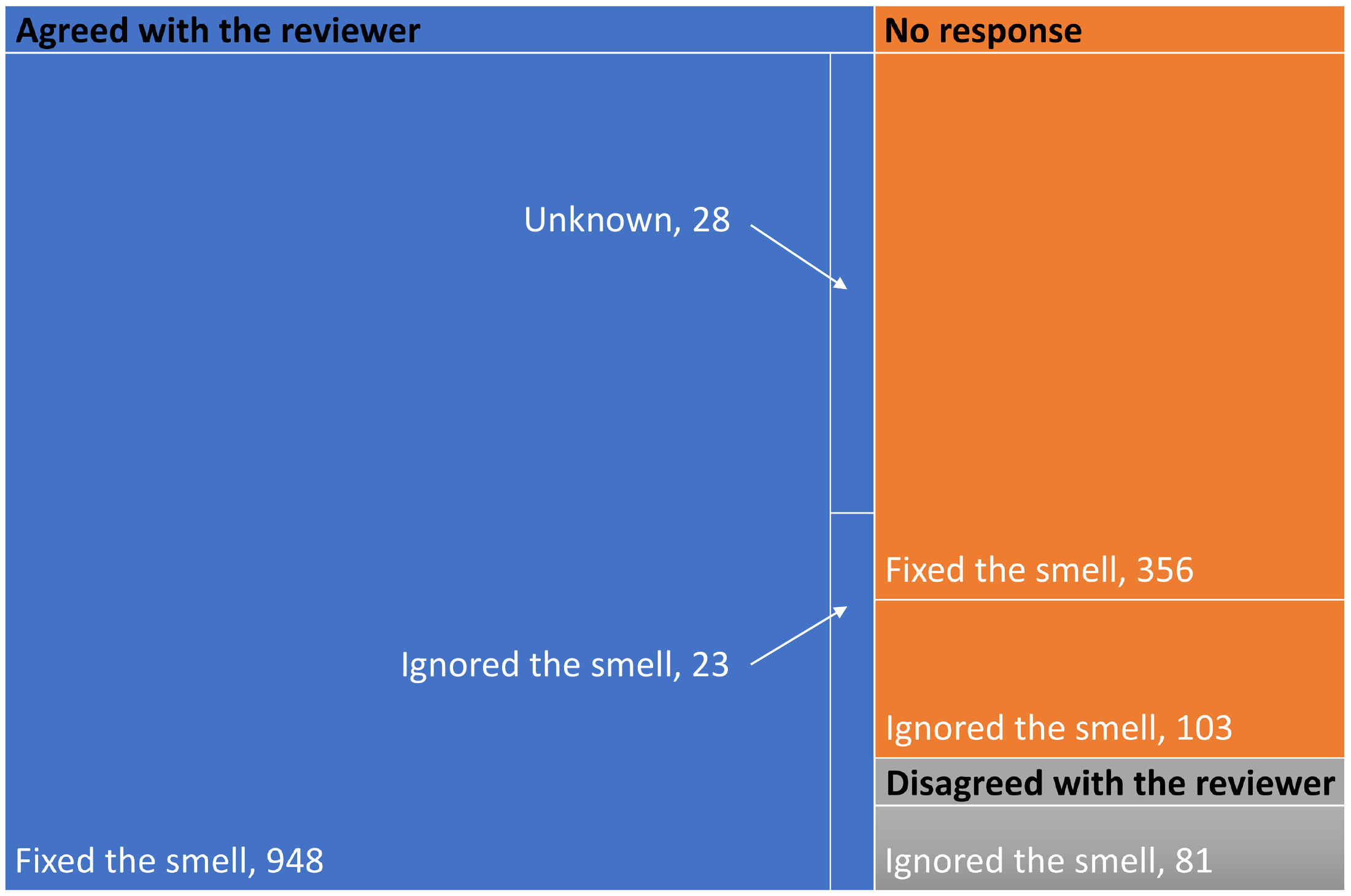}
    \caption{A treemap of the relationship between developers' actions in response to reviewers' recommendations regarding code smells identified in the code}
    \label{fig:actions_map}
\end{figure}

For answering this RQ, a visual map of reviewer recommendations and resulting developer actions is shown in Fig. \ref{fig:actions_map}.

In 971 (63\%) of the obtained reviews, developers agreed with the reviewer's suggestion and took exactly the same actions (either \textit{fix} or \textit{ignore}) as suggested by the reviewers. Of those cases, there are 23 cases where developers agreed with reviewers on ignoring the smells (i.e., a smell had been identified, but the reviewer may have thought that the impact of the smell was minor and \textit{there is no need to fix the smell now}). The example below shows a case where a reviewer pointed out that they could accept duplicated code if there was a reasonable justification and the developer gave their explanation and ignored the smell.

\begin{qoutebox}{white}{}
\textbf{Link:} \url{http://alturl.com/s59so}\\
\textbf{Reviewer:} ``...I just don't like duplicated code but if there is a reasonable justification for this I can be sold cheaply and easily.'' \\
\textbf{Developer:} ``we need \texttt{create\_vm} here to support a lot of the other testing in this method. I agree it's duplicate code, but it's needed here too and this one is more complex that (sic) the \texttt{test\_config} one....''
\end{qoutebox}

There are 28 (2\%) reviews where the developers agreed with reviewers but did not make any changes to the code immediately; however, the developers promised to fix the identified smells in a later patch or commit. Below is an example in which the developer promised to remove the duplication in a follow-up change.

\begin{qoutebox}{white}{http://alturl.com/ziks2}
\textbf{Link:} \url{http://alturl.com/pzmzz}\\
\textbf{Reviewer:} ``Can't we use the same enum class for both instead of keeping in sync? (for example in follow-up patch)''\\
\textbf{Developer:} ``Yes, I will remove the duplication in a follow up change.''
\end{qoutebox}

In 356 (23\%) reviews, even when developers did not respond to reviewers directly in the review system, they still made the required changes to the source code files. We note that there are other 81 (5\%) reviews where developers had different opinions from reviewers and decided to ignore the recommendations to refactor the code and remove the smell. In those cases, the developers themselves decided that the smell was either not as critical as perceived by the reviewers, or there were time or project constraints preventing them from implementing the changes, which are typically self-admitted technical debt \citep{potdar2014exploratory}. An example review is shown below:

\begin{qoutebox}{white}{}
\textbf{Link:} \url{http://alturl.com/pzmzz}\\
\textbf{Reviewer:} ``This method has a lot duplicated code of `\_apply\_instance\_name\_template'. The differ in the use of `index' and the CONF parameters. With a bit refactoring only one method would be necessary I guess.''\\
\textbf{Developer:} ``I thought to make / leave this separate in case one wants to configure the multi\_instance\_name\_template different to that of single instance.''
\end{qoutebox}

Similarly, there are also 103 (7\%) reviews in which developers neither replied to reviewers nor modified the source code. For those cases, we suppose that the developers did not find the recommendations regarding how to deal with the specific smells in the code helpful and therefore decided not to perform any changes. In all of those cases, no further explanation/reasons were provided by the developers on why they ignored these recommended changes.\\ 

\noindent\resizebox{\columnwidth}{!}{\fbox{
	\parbox{\columnwidth}{
	\textbf{RQ3 Summary:} In most reviews, reviewers provide fixing (refactoring) recommendations (e.g., in the form of code snippets) to help developers remove the identified smells. Developers generally follow those recommendations and perform the suggested refactoring operations, which then appear in the patches committed after the review.
	}
}}

\subsection{RQ4: How long does it take to resolve code smells by developers after they have been identified by reviewers?}

According to the result of RQ3.2, a total of 1,304 (85\%) code smells identified in the reviews were fixed by developers. Of these, we removed cases where the time taken for the fix was more than one year (4 reviews), the resolution time could not be determined (2 reviews), and the identification time was earlier than the resolution time (2 reviews, as explained in Section \ref{methodology_for_rq4}). We also note that some smell categories are very few in number from the 1,304 code reviews (e.g., \textit{long method} and \textit{circular dependency}) and we chose to exclude them (35 reviews) from the analysis of this RQ because their resolution time may have limited statistical significance. Thus, the main smell categories we investigated in this RQ are \textit{duplicated code}, \textit{bad naming}, and \textit{dead code} smells. Finally, we analysed 1,261 code reviews to answer RQ4. Table \ref{tab:time_taken_for_fixing_smells} shows minimum, quartile, maximum, and mean time for fixing different categories of code smells.

\begin{table}[h]
\caption{Time taken for fixing different categories of code smells}
\label{tab:time_taken_for_fixing_smells}
\resizebox{\columnwidth}{!}{
\begin{tabular}{@{}l|c|c|c@{}}
\toprule
\textbf{}               & \textbf{Duplicated Code}      & \textbf{Bad Naming}       & \textbf{Dead Code} \\ \midrule
Count                   & 555                           & 401                       & 305       \\
Minimum Time (second)   & 69s                           & 127s                      & 63s       \\
Lower Quartile (hour)   & 7.9h                          & 3.4h                      & 2.1h      \\
Median Time (hour)      & 23.0h                         & 20.5h                     & 19.1h     \\
Higher Quartile (day)   & 5.0d                          & 3.8d                      & 3.7d      \\
Maximum Time (day)      & 352.0d                        & 325.3d                    & 291.2d    \\
Mean Time (day)         & 12.2d                         & 9.2d                      & 7.8d      \\  \bottomrule
\end{tabular}}
\end{table}

\begin{figure}[htb]
    \centering
    \captionsetup{justification=centering}
    \includegraphics[width=1\linewidth]{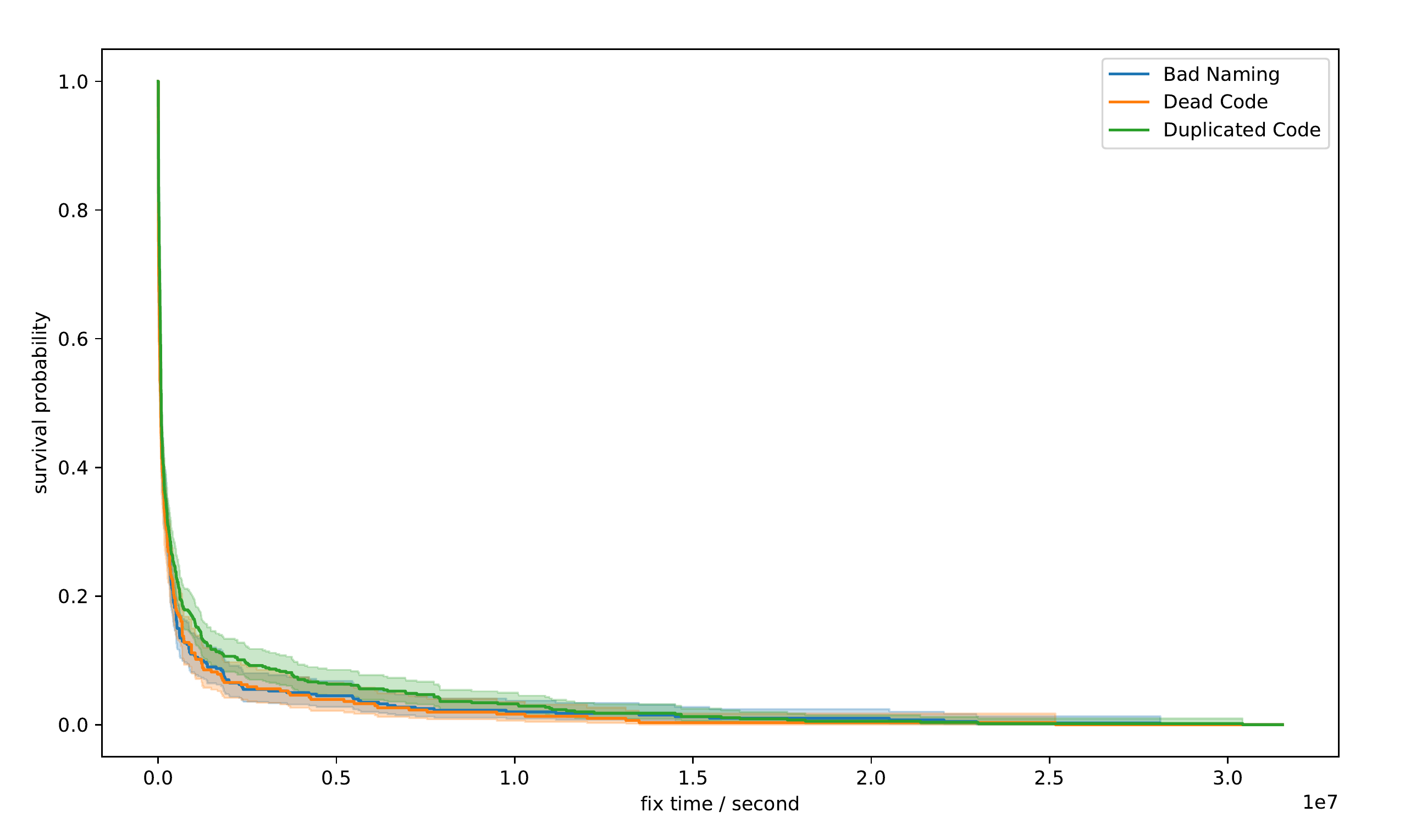}
    \caption{The survival curve of different smells identified in code reviews}
    \label{fig:survival_curve}
\end{figure}

As shown in Table \ref{tab:time_taken_for_fixing_smells}, we found that the minimum time taken for fixing \textit{duplicated code}, \textit{bad naming}, and \textit{dead code} could be very short - only around one or two minutes. However, it may also take a very long time to fix these code smells (around a year in some cases). According to the quartile time (i.e., lower quartile, median, and higher quartile time) for fixing these three categories of code smells, \textit{duplicated code} was fixed slower than the other two smells. We also conducted a survival analysis using the Kaplan–Meier curves \citep{kaplan1958nonparametric}. The survival curve is shown in Fig \ref{fig:survival_curve}, which also supports our finding above.

Further, we used a Kruskal-Wallis Test \citep{kruskal1952use} to determine whether or not there was a statistically significant difference in fix time between these three groups. We used SciPy\footnote{\url{https://scipy.org/}} (fundamental algorithms for scientific computing in Python) to perform the Kruskal-Wallis Test. In this case, the test statistic is 13.3041 and the corresponding \textit{p-value} is 0.0013. Since the \textit{p-value} is less than 0.05, we can conclude that the category of code smell lead to a statistically significant difference in fix time.

\begin{figure}[htb]
    \centering
    \captionsetup{justification=centering}
    \includegraphics[width=1\linewidth]{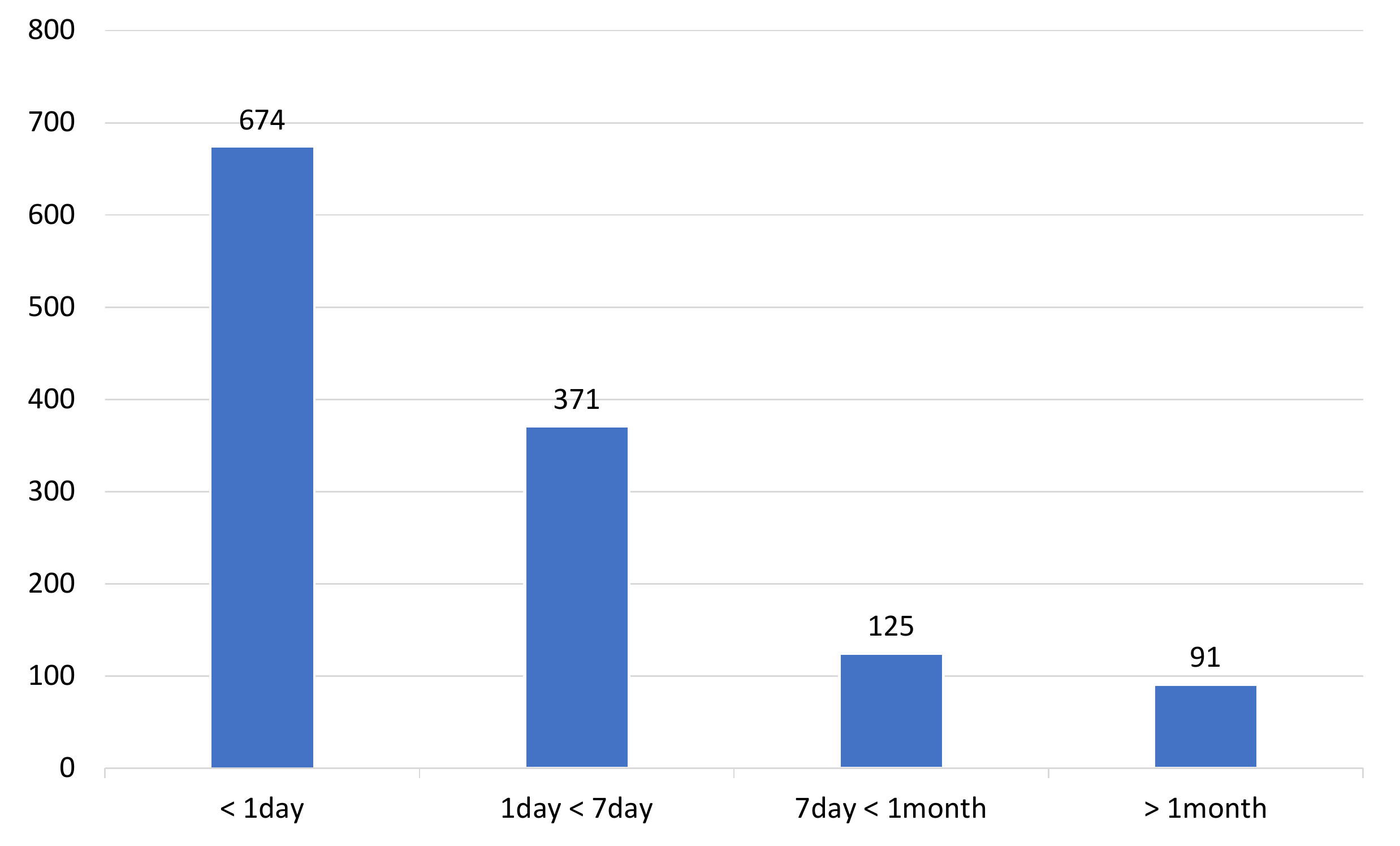}
    \caption{Distribution of time for fixing smells identified in code reviews}
    \label{fig:distribution_of_fixing_time}
\end{figure}

Fig. \ref{fig:distribution_of_fixing_time} shows the distribution of time taken for fixing those code smells. From this figure, we observe that 1,045 smells (83\%) were fixed within one week. More than half of smells (674, 53\%) were fixed within one day, and 371 (29\%) smells were fixed in more than one day but within one week. 125 (10\%) smells took developers 2-4 weeks to fix, and only 91 (7\%) smells took more than one month to fix.

We also calculated the minimum, quartile, maximum, and mean time for fixing code smells based on whether the specific refactoring actions were provided or not. The results are shown in Table \ref{tab:time_taken_for_fixing_smells_by_action}. We can see that the time taken to fix the identified smells is almost the same under these two circumstances. We formed a null hypothesis that there was no statistical difference in time taken to fix a smell whether specific refactoring suggestions have been provided or not. We then performed a Mann-Whitney Test \citep{mann1947test} to compare the differences between those two groups. The test statistic is 160548.5 and the corresponding \textit{p-value} is 0.7800 (i.e., \textit{p-value}\textgreater 0.05); it can therefore be concluded that there is no statistically significant difference between fix time in these two categories. Based on the results, we can see that whether reviewers provided specific refactoring suggestions or not has little effect on the fix time of the identified code smells.\\

\begin{table}[h]
\caption{Time taken for fixing smells (classification according to whether the reviewers provide specific refactoring actions or not)}
\label{tab:time_taken_for_fixing_smells_by_action}
\resizebox{\columnwidth}{!}{
\begin{tabular}{@{}l|c|c@{}}
\toprule
\textbf{}               & \textbf{Specific refactoring actions provided}    & \textbf{No specific actions provided} \\ \midrule
Count                   & 360                                               & 901       \\
Minimum Time (second)   & 69s                                               & 63s      \\
Lower Quartile (hour)   & 4.5h                                              & 4.6h      \\
Median Time (hour)      & 21.6h                                             & 21.1h     \\
Higher Quartile (day)   & 3.9d                                              & 4.2d      \\
Maximum Time (day)      & 266.1d                                            & 325.0d    \\
Mean Time (day)         & 8.2d                                              & 10.9d     \\  \bottomrule
\end{tabular}}
\end{table}

\noindent\resizebox{\columnwidth}{!}{\fbox{
	\parbox{\columnwidth}{
		\textbf{RQ4 Summary:} Among the studied smells, \textit{duplicated code} smells took more time to fix compared to \textit{bad naming} and \textit{dead code} smells. Moreover, 83\% of the smells were fixed by developers within one week from being identified in code reviews.
	}
}}

\subsection{RQ5: What are the common causes for not resolving code smells that have been identified in code?}

\begin{figure}[htb]
    \centering
    \captionsetup{justification=centering}
    \includegraphics[width=1\linewidth]{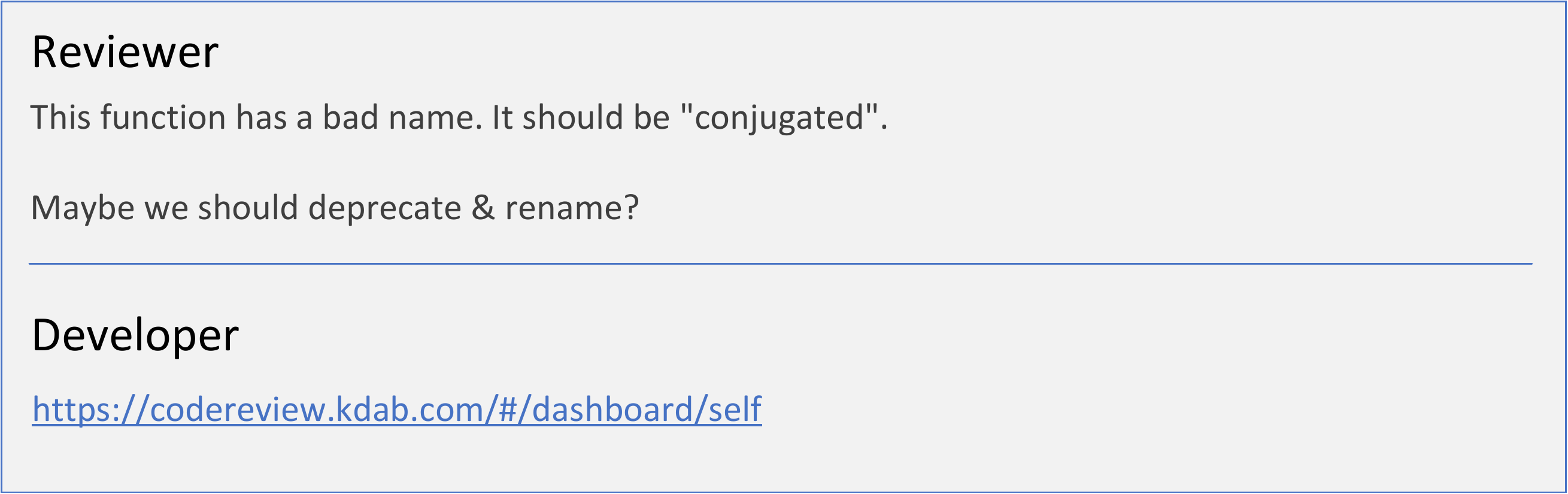}
    \caption{The review in which we could not find the reason why the developer ignored the smell}
    \label{unknown_reason_example}
\end{figure}

According to the results of RQ3.2, there are 81 (5\%) reviews in which developers disagreed with reviewers and chose to ignore the identified code smells. 
We excluded one review as we were not able to access the URL link provided by the developer. Fig. \ref{unknown_reason_example} shows this review, in which the developer only replied with a URL to the reviewer, but to which we had no access. We then inspected the later patches and the final status of the code change. We found that the developer made no change in later patches and the code change was finally merged to the code base. Considering this, we treated this as a case of an \textit{ignore} of the reviewer's suggestion. We could not find the reason why the developer ignored the smell and consequently we excluded this review from our analysis. 

\begin{table}[h]
\caption{Distribution of causes for ignoring smells identified in code reviews}
\label{tab:causes_for_ignoring_smells}
\resizebox{\columnwidth}{!}{
\begin{tabular}{@{}l|c|c@{}}
\toprule
\textbf{The cause}                                      & \textbf{Count}    & \textbf{\%}      \\ \midrule
Not worth fixing the smell                              & 28                & 35\%                  \\
Difference in opinion between developers and reviewers  & 20                & 25\%                  \\
Limited by developers' knowledge                        & 9                 & 11\%                  \\  
Keep consistent with other code                         & 8                 & 10\%                  \\
For future consideration                                & 7                 & 9\%                   \\
Cause other errors when fixing the smell                & 4                 & 5\%                   \\
Improve code readability                                & 4                 & 5\%                   \\
\bottomrule
\end{tabular}}
\end{table}

As shown in Table \ref{tab:causes_for_ignoring_smells}, we found that the main reason for why developers ignored identified smells was simply that it was \textbf{not worth fixing the smell} (in 28 reviews). In this case, the developers thought that there were more important things to consider, or fixing the smells would bring little value or add more complexity. An example of such a case is shown below. In this case, the developer thought that removing the duplication would make specific comparisons more expensive and the developer chose to ignore the smell.

\begin{qoutebox}{white}{}
\textbf{Link:} \url{http://alturl.com/76z9f} \\
\textbf{Reviewer:} ``How about implementing int compare() which returns \textless 0, 0 or \textgreater0, and then implement the other comparisons in terms of that? Should result in less code duplication as there currently is with operator\textless and operator==
...''\\
\textbf{Developer:} ``I don't see how this reduces code duplication. I can see how a compare would allow one function to do all comparisons but it would make specific comparisons more expensive.''
\end{qoutebox}

20 reviews attribute \textbf{difference in opinion between developers and reviewers} to be the reason why developers ignored the smells. This means that although the reviewers identified the code smell, the developer thought it was acceptable and chose not to fix it. 
In the below example, the reviewer suggested removing the \textit{duplicated code} while the developer thought that there was no problem and ignored the smell.

\begin{qoutebox}{white}{}
\textbf{Link:} \url{http://alturl.com/up6a5} \\
\textbf{Reviewer:} ``It's kind of crazy to have to duplicate this kind of logic in the tests.  The first thing I'd like to suggest is extracting event name determination into its own method so it can be easily mocked out in unrelated tests.''\\
\textbf{Developer:} ``I'd prefer to just leave this for now if that's okay''
\end{qoutebox}

In nine reviews, the reason why developers ignored the identified smell was \textbf{limited by developers' knowledge}. This means that developers could not find a better way to fix the identified smell or they did not have enough information to fix it. Here is an example about \textit{bad naming} where the developer could not find a better name and left it as it was.

\begin{qoutebox}{white}{}
\textbf{Link:} \url{http://alturl.com/z753q} \\
\textbf{Reviewer:} ``naming might be confusing. You could also include Result::MessageIntermediate and think of a better name for this group... (just to save an unnecessary roundtrip: no suggestions.)'' \\
\textbf{Developer:} ``got no better naming idea as well.. leaving this for now as is, guess this will change when the whole Message* stuff gets re-done.''
\end{qoutebox}

In eight reviews, the reason why developers ignored the smell is that they chose to \textbf{keep consistent with other code}. The following example shows that the developer just used the original name, although the reviewer thought that it was poor naming.

\begin{qoutebox}{white}{}
\textbf{Link:} \url{http://alturl.com/onn6f} \\
\textbf{Reviewer:} ``Gosh, what an awful naming convention...'' \\
\textbf{Developer:} ``I agree but I just used the original one''
\end{qoutebox}

Seven reviews indicate that developers chose to ignore smells \textbf{for future consideration}. Developers indicated that they preferred to keep the status quo to help future changes. There are also cases in which developers indicated that the smells would be fixed once some certain features came online. Below is an example in which the developer believed that small duplication would help to make things cleaner for subsequent changes.

\begin{qoutebox}{white}{}
\textbf{Link:} \url{http://alturl.com/bevos} \\
\textbf{Reviewer:} ``This really needs to be extracted into a common method. We will suck at maintaining the API if we have this level of duplication.'' \\
\textbf{Developer:} ``I think we could remove duplication for force param here, but leave things as is for other params, e.g. common get\_args method will add microversion checks for block\_migration param, see lines 83-97, which I'd like to leave for microversion 2.34, because it's already supports 'auto' for block\_migration. So this small dup will help to make things cleaner for subsequent changes.''
\end{qoutebox}

Four reviews indicated that dealing with a particular smell could \textbf{cause other errors when fixing the smell}. The following example shows that removing the identified \textit{dead code} (i.e., the \texttt{return} statement) would produce errors in release mode.

\begin{qoutebox}{white}{}
\textbf{Link:} \url{http://alturl.com/au5pu} \\
\textbf{Reviewer:} ``No dead code, please.'' \\
\textbf{Developer:} ``It's dead code but if I omit the "return 0" statement I get errors in release mode about missing return value.''
\end{qoutebox}

In the remaining 4 reviews, the developers noted that the existence of smells would have a positive impact and it would, in general, \textbf{improve code readability}, as shown in the example below.

\begin{qoutebox}{white}{}
\textbf{Link:} \url{http://alturl.com/uzo6c}\\
\textbf{Reviewer:} ``nit: duplicates to lines 444-449; could be refactored into an attribute.'' \\
\textbf{Developer:} ``True, but that would make the test a little less readable IMO. In this case I think the duplication is worth it''
\end{qoutebox}

We also checked the status of the code changes where developers disagreed with reviewers and chose to ignore the identified smells. The detailed results are show in Table \ref{tab:code_change_status}. Although the identified smells were ignored by developers, 61 (75.3\%) code changes were still merged to the primary codebase while only 20 (24.7\%) code changes were finally abandoned or deferred.

\begin{table}[h]
    \centering
    \caption{The status of code change where developers disagreed with reviewers and ignored the smells}
    \label{tab:code_change_status}
    \begin{tabular}{c c c}
        \toprule
        \makebox[0.3\columnwidth][c]{\textbf{Code Change Status}}     & \makebox[0.2\columnwidth][c]{\textbf{Count}}        & \makebox[0.2\columnwidth][c]{\textbf{\%}}      \\
        \midrule
        Merged                          & 61                    & 75.3\%                \\
        Abandoned                       & 19                    & 23.5\%                \\
        Deferred                        & 1                     & 1.2\%                 \\
        \bottomrule
    \end{tabular}
\end{table}

\noindent\resizebox{\columnwidth}{!}{\fbox{
	\parbox{\columnwidth}{
		\textbf{RQ5 Summary:} Developers disagreed with reviewers in only a small number of code reviews (81, 5\%). In terms of the reasons for disagreement, \textbf{not worth fixing the smell} was the main reason for ignoring the identified smells. Although developers disagreed with reviewers and ignored the smells, 75.3\% of these changes were still merged into the codebase.
	}
}}
\section{Discussion}
\label{sec:discussion}

\subsection{RQ1: The most frequently identified smells}
\label{sec:discussion:rq1}

In general, most of the smells are not extracted from the code review data. One potential reason is that code smells do not appear frequently during the development of the four selected projects in this study, or, simply, that the reviewers were unaware of the presence of certain code smells. Another potential reason is that reviewers were aware of code smells, but that they did not consider them very harmful. One avenue of further work would be to run smell detection tools and compare the smells identified by these tools with the smells identified manually by reviewers to better understand how many code smells are being missed by reviewers. Interviews with the code reviewers can also provide further understanding of the reasons behind the low number of code smells being discussed in code review. 

The results of RQ1 imply that \textit{duplicated code}, \textit{bad naming} and \textit{dead code} are, by far, the most frequently identified code smells in code reviews. Results regarding \textit{duplicated code} are in line with previous findings which indicate that this smell is frequently discussed among developers in online forums \citep{tahir2020stackexchange} and is also the smell that developers are most concerned about \citep{Yamashita2013e}. However, \textit{dead code} and \textit{bad naming} were not found to be ranked highly in previous studies \citep{Yamashita2013e}. The different results are due to the different context and domain, critical to identifying smells, as shown by previous studies \citep{Yamashita2013e,tahir2020stackexchange}. The results reported in these two previous studies \citep{tahir2020stackexchange, Yamashita2013e} are based on a more generic investigation of code smells among online Q\&A forum users and developers. The context of some of these code smells is not fully taken into account, even if the developers provide specific scenarios to explain their views. In contrast, the results reported in this study are project-centric and the context of the identified code smells during code reviews is known to reviewers and developers involved in the identification and removal of the smells. This is also supported by our further investigation of the refactoring actions applied to source code once smells are identified (discussed in RQ3 and RQ5).

A study by \cite{palomba2018@maintainability} has shown that the most diffuse code smells are those characterized by long and/or complex code (e.g., \emph{complex class}), which is different from our results of RQ1. One potential reason for this is the contrast in code smells considered in the two studies. In the study of \cite{palomba2018@maintainability}, they did not consider \emph{duplicated code}, \emph{dead code} and \emph{bad naming}, but focused on a larger granularity of code smells, i.e., at the class and method level. Another potential reason is that, in modern code review, code changes are kept as small as possible to facilitate the review process and this may contain very few class level design issues, e.g., \emph{complex class}. Code smells at a lower level of granularity (e.g., \emph{dead code}) are easier to detect on the fly (especially with conditional statements), while complexity related smells can be hard to detect without the use of detection tools. The reasons for the difference are also an interesting aspect that should be explored in future research.

In addition, different styles of code reviews may affect smell detection. In this work, we focused on the modern code review process that reviews code changes. Compared with other styles of code reviews (e.g., reviewing code instead of changes to the code base), changes by default are going to be smaller in a modern code review setting, leading to detection of smells with a smaller granularity. Other styles of code review may consider the whole project and it is more likely to identify smells at higher levels, such as project and component level code smells.

\subsection{RQ2: The causes for identified smells}
In general, we identified four types of common causes (see Fig. \ref{fig:reasons_for_smells}) for code smells in code reviews (RQ2). Among these, \textbf{violation of coding conventions} was the major cause of code smells identified in reviews. Conventions are important in reducing the cost of software maintenance, while the existence of smells can increase this cost. We conjecture that this is because developers may not be familiar with the coding conventions of their community and the system they implemented. More specifically, \textbf{violation of coding conventions} is the main cause for the \textit{bad naming} smell. Usually, communities or companies will have a specific naming convention, which can help improve the readability of code. 
The main cause for \textit{duplicated code} and \textit{dead code} is \textbf{lack of familiarity with existing code}. For example, \textit{duplicated code} and \textit{dead code} may occur because developers are unaware of existing functionality. This implies that a developer's unfamiliarity with coding conventions or existing code can inadvertently lead to smells or other problems and this can have a negative impact on software quality.

Another main observation is that more than half of reviewers (in review comments where they indicated that there was a code smell) simply pointed out the smells in the code, but did not provide any further explanation as to why they considered that as a smell. One explanation for this is that the identified smells are simple or self-explanatory (e.g., \textit{duplicated code}, \textit{dead code}). Therefore, it is not expected that the reviewers needed to provide any further explanation for these smells. Although the point of code review is to identify shortcomings (including potential code smells) in contributed code, understanding the causes of code smells can help practitioners better understand how the code smell was introduced and then take corresponding remedial measures. 

\subsection{RQ3: The relationship between what reviewers suggest and the actions taken by developers}
\label{discussion_of_rq3}
The results of RQ3 show that reviewers usually provide useful recommendations (sometimes in the form of code snippets) when they identify smells in the code and developers usually follow these suggestions. Given the constructive nature of most reviews, developers tend to agree with the review-based smell detection mechanism (i.e., where a reviewer detects and reports a smell) and, in most cases, they perform the recommended actions (i.e., refactoring their code) to remove the smell. We believe that this is because reviewers can take more information into account as the program context and domain are important in identifying smells \citep{Yamashita2013e,tahir2020stackexchange,sae2018context}.

The result of RQ3.1 shows that reviewers usually provide general refactoring instructions (i.e., remove or refactor the smells) without specific suggestions (i.e., how to refactor the smells) and these types of smells are usually easy to fix. For example, \textit{dead code} is usually resolved by simply removing the unused or unreachable code. For \textit{bad naming}, it is usually a matter of coming up with a different name and changing a small part of source code. Moreover, compared with the recommendations for \textit{dead code} and \textit{bad naming}, there are more types of specific refactoring actions suggested by reviewers for resolving \textit{duplicated code}. One possible reason could be that \textit{duplicated code} is more difficult to repair by contrast and consequently reviewers would propose more detailed actions to help developers remove those smells. Another finding is that \textbf{Extract Method} is the most frequently suggested refactoring action, which is consistent with what Fowler states in his seminal refactoring text \citep{martin1999refactoring}: ``\textit{Extract Method is one of the most common refactorings I do. I look at a method that is too long or look at code that needs a comment to understand its purpose. I then turn that fragment of code into its own method}''. 

There were some case when reviewers suggested ignoring identified smells. In these cases, we found that \textit{duplicated code} made up the majority. In general, the tolerance of \textit{duplicated code} varies from one reviewer to another. When reviewers identified \textit{duplicated code}, but the number of lines of duplicated code did not reach their threshold, reviewers often indicated that the relevant \textit{duplicated code} could be ignored.

\subsection{RQ4: The time taken for fixing the smells}
From the perspective of code smell categories, it usually takes more time to fix \textit{duplicated code} than \textit{dead code} and \textit{bad naming}. We believe that this is related to the nature of those code smells. Usually, code duplication involves multiple parts of the source code rather than a single part, which makes it more difficult to fix than \textit{dead code} and \textit{bad naming}.
Another finding is that the longest time taken for fixing \textit{duplicated code}, \textit{bad naming}, or \textit{dead code} was around 300 days. We posit that this is partially related to the way developers work on patches and abandon code changes. When the developer uploads a new patch, it may not be used to solve the identified code smell and to solve other problems also. In some reviews, we regarded the time of abandoning the code change where the smell locates as the resolution time of the identified code smell. This could also prolong the resolution time of the code smell we obtained because the code change is usually abandoned after a long time without any update.

Moreover, from the perspective of the distribution of the time taken to fix smells, most of the identified smells were fixed within one week from the time they were first identified in the reviews. We suspect that this finding may be related to the nature of the code smell and the reviewers' recommendations. \textit{Dead code} and \textit{bad naming} are usually easier to fix as we explained in Section \ref{discussion_of_rq3}. Additionally, we found that 7\% of smells took more than one month to fix. We then further checked related information on these smells, i.e., the code review discussions, but found that no reasons were provided for such a delay in most cases. The developers just uploaded the patches or abandoned the code changes after a significant time period without providing any reasoning. Only in one review\footnote{\url{http://alturl.com/rrxo7}}, a developer indicate that it was not the right time to fix the identified code smell.

\subsection{RQ5: Reasons for ignoring the identified smells}
Although not as frequently occurring, there are cases where changes recommended by reviewers were ignored (see Figure~\ref{fig:actions_map}). The result for RQ5 shows that \textbf{not worth fixing the smell} is the main reason why developers ignored removing the smells from the code. In other words, it is assumed that fixing the smells would either add more complexity to the code, or just bring little value as a result. The context of the smell (such as the time of fixing it, the complexity it brings, whether there is something more important, etc.) should be taken into full consideration and the value that the fix will bring should also be assessed.

Another reason is \textbf{difference in opinion between developers and reviewers},  consistent with what Fowler noted \citep{martin1999refactoring} that ``\textit{no set of metrics rivals informed human intuition''}. This situation is partially due to the different understanding or experience of reviewers and developers about the severity of identified code smells. When a reviewer identifies a code smell to be resolved, a developer may not agree that the code smell needs to be fixed; equally,  that it is an issue which can be fixed later in the same way that technical debt is accrued \citep{li2015systematic}.

We also found that although developers ignored smells, most of the code changes were still merged into the codebase. One potential reason for this is that the proposal of a code change is not specifically used to fix the identified code smell. The introduction of a code smell is usually a side product and the existence of a code smell may have little influence on the merging of code change. It also means that the program context of a code smell has a great impact on how harmful the smell is and whether or not it needs to be fixed immediately.

\subsection{Implications}
There are a number of implications of the work contained in this paper. 
First, although we built the initial set of keywords with 5 general code smell terms and 40 specific code smell terms, most of the smells are not extracted from the code review data (e.g., \textit{long parameter list}, \textit{temporary field}, and \textit{lazy class}). Gerrit is designed to review code changes and smells might not be mentioned during the code review process if they are not deemed severe enough by the reviewers. Another potential reason is that code smells considered as problematic in academic research may not be considered as a pressing problem in industry. Thus, more research should be conducted with practitioners to explore existing code smells and to understand the driving force behind industry efforts on code smell detection and elimination. This will help to guide the design of next-generation code smell detection tools.

Second, \textbf{violation of coding conventions} is the main cause of code smells identified in code reviews. It implies that a developer's lack of familiarity with the coding conventions in their company or organization could have a significantly negative impact on software quality. To reduce code smells, project leaders need to adopt code analysis tools and also help educate their developers to become familiar with the coding conventions adopted in the system; we note that some tools can also be used to automatically check for compliance with code conventions. 

Third, in smell-related reviews, reviewers usually give useful suggestions to help developers better fix the identified code smells and developers generally tend to accept those suggestions. Review-based detection of smells is seen as a trustworthy mechanism by developers. Although code analysis tools (both static analyzers and dynamic (coverage-based) tools) are able to find some of those smells, their large outputs restrict their usefulness. Most tools are context and domain-insensitive, making their results less useful due to potential false positives produced by these tools \citep{Fontana2016}. 

Fourth, it usually takes developers less than one week to fix an identified smell. According to the results of RQ4, providing detailed recommendations has little influence on the fix time of code smells. Fixing those smells depends on many factors, most importantly program context. For relatively less complex code smells (e.g., \textit{duplicated code}), reviewers may merely point out the existence of those smells rather than spending time making more detailed suggestions for their removal.

Fifth, there are cases where developers disagreed with reviewers and ignored identified smells (see Figure~\ref{fig:actions_map}). Of these, \textbf{not worth fixing the smell} is the main cause when developers chose to ignore identified smells. This could imply that developers do not tend to make any changes to existing code where fixing code smells takes significant effort (a typical technical debt scenario \citep{li2015systematic}.
That is, context seems to matter in deciding whether a smell is bad or not \citep{tahir2020stackexchange,sharma2018survey}. 
There have been some recent attempts to develop smell-detection tools that take developers-context into account \citep{sae2018context,pecorelli2020developers}. However, contextual factors such as project structure and developer experience are much harder to capture with tools. Code reviewers are much better positioned to understand and account for those contextual factors (as they are involved in the project) and their assessment of smells might be trusted more by developers than that of automated detection tools.

Finally, to increase the reliability of detecting code smells, it may need a two-step detection mechanism; first, static analysis tools to identify smells (as they are faster than human assessment and also more scalable) and second for reviewers to go through those smell instances. They should then decide, based on the additional contextual factors, which of those smells should be removed and at what cost. One potential problem with such an approach is that most tools would probably produce large sets of outputs, making it impractical for reviewers working on a large code base; improving the accuracy of smell-detection tools is vital for its application in such a context.
\section{Threats to Validity}
\label{sec:threats}

Given the empirical nature of our study, potential threats can affect the study results. We classify and discuss these threats by following the recommendations suggested in \cite{Wohlin2000Experimentation}.

\textbf{External Validity}: Our study considered two major projects from the OpenStack community (Nova and Neutron) and two major projects from the Qt community (Qt Base and Qt Creator), since those projects have invested a significant effort in their code review process (see Section \ref{sec:research_setting}). OpenStack is a set of software tools for building and managing cloud computing platforms and Qt is an open source cross-platform application and UI framework. The projects from the OpenStack community are mainly written in Python, while the projects from the Qt community are mainly written in C++. Different domains and programming languages can help improve the external validity and make the study results and findings more generalizable to other systems. We believe that our results and findings could help researchers and developers understand the importance of the manual detection of code smells better. Moreover, including code review discussions from other communities will supplement our findings and this may lead to more general conclusions.

\textbf{Internal Validity}: The main threat to internal validity is related to the quality of the selected projects. It is possible that the projects we selected do not provide a good representation of the types of code smells we included in our study. To address this threat, we selected two large projects from the OpenStack community and two large projects from the Qt community with Gerrit as their code review tool. Their investment in code review processes and commitment to perform code review to their entire code base make them good candidates for our study. Another threat to the internal validity is the nature of modern code review used in Gerrit. In such code review platforms, code changes (instead of a code snapshot or the whole codebase) are reviewed, where the changes are usually micro to small changes. The practice with such a review system is to review small additions or modifications to the codebase. By default, this will limit higher, more abstract project-level smells and issues from being detected.

\textbf{Construct Validity}: A large part of the study depends on manual analysis of the data, which could affect the construct validity due to personal oversight and bias. In order to reduce its impact, each step in the manual analysis (i.e., identifying smell-related code reviews and their classifications in various aspects) was conducted by at least two authors and a third author was involved in case of disagreement. The selection of the keywords used to identify the reviews which contain smell discussions is another threat to construct validity since reviewers and developers may use terms other than those that we used in our mining query. To minimize the impact of this threat, we first combined a list of code smell terms that developers and researchers frequently used, as reported in several previous studies \citep{Tahir2018EASE, martin1999refactoring, zhang2011code}. Then, we identified the keywords by following the systematic approach used by \cite{bosu2014identifying} to minimize the impact of missing keywords due to misspelling or other textual issues. Moreover, we randomly selected a collection of code review comments that did not contain any of our keywords to supplement our dataset, reducing the threat to construct validity.

\textbf{Reliability}: Before starting our full scale study, we conducted a pilot run to check the suitability of the data source. The execution of all the steps in our study, including the process of data mining, data filtering and manual analysis was discussed and confirmed by at least two of the authors. We also provided the replication package of this work online \citep{anonymous_replication_package} for replication purposes, which partially increases the reliability of the study results.

\section{Conclusions}
\label{sec:conclusion}

In this work, we conducted an empirical analysis of code smells identified in modern code review (MCR). Although there are many studies focusing on code smells or code reviews, little is known about the extent to which code smells are identified and resolved during MCR. To this end, we statistically analysed the code review comments from four most active projects of the OpenStack community (Nova and Neutron) and the Qt community (Qt Base and Qt Creator). More specifically, we manually analysed the \textit{types}, \textit{causes}, \textit{actions}, and \textit{fixing interval} of/towards the identified code smells.

According to our results, code smells are not commonly identified in code reviews and when identified, \emph{duplicated code}, \emph{bad naming} and \emph{dead code} are, by far, the most frequently identified smells. When smells are identified, most reviewers provide constructive suggestions to help developers fix the code and developers are willing to fix the smells through suggested refactoring operations; it usually takes developers less than one week to fix the identified smells.
We also found that code smells were often introduced as a result of developers violating coding conventions. Although not as frequent, there are also cases where developers disagree with the reviewers and ignore identified smells. The main cause for it is that developers think it is not worth fixing the smells (i.e., bring little value or introduce more complexity when fixing).

Based on our findings, we make the following suggestions for both researchers and practitioners:

\begin{enumerate}
    \item Developers should follow the coding conventions in their projects to reduce code smell incidents; some tools can also be used to automatically check for compliance with code conventions.
    \item Code smell detection via code reviews is seen as a trustworthy approach by developers (given their constructive nature) and smell-removal recommendations made by reviewers appear more actionable by developers.
    \item Program context is important in the identification of code smells and also should be taken into account in order to determine whether to fix the identified code smells immediately.
\end{enumerate}

In the next step, we plan to extend this work by studying code smells in code reviews in a larger set of projects from different communities, including from industrial projects. We also plan to obtain a further understanding of the practitioners' attitude towards code smells in code reviews by conducting a survey with both reviewers and developers and thus take a closer look at what the academic and industrial communities think about code smells via code review.

\section*{Acknowledgements}
This work is supported by the National Natural Science Foundation of China (NSFC) under Grant No. 62172311 and the Special Fund of Hubei Luojia Laboratory.

\bibliographystyle{spbasic}
\bibliography{references}

\end{document}